\documentclass[12pt]{article}
\usepackage{fancyhdr}

\usepackage{mathrsfs}
\usepackage[T1]{fontenc}
\usepackage{mathpazo}
\usepackage{setspace}
\usepackage{amsfonts}
\usepackage{amssymb}
\usepackage{amsmath}
\usepackage{epsfig}
\usepackage{latexsym}
\usepackage{color}
\usepackage{graphicx}
\usepackage{nicefrac}
\usepackage[latin1]{inputenc}
\usepackage{slashed}
\usepackage{multirow}
\usepackage{comment}
\usepackage{soul}
\usepackage{hyperref}
\usepackage[nosort]{cite}
\usepackage{datetime}

\usepackage{braket}

\usepackage[titletoc]{appendix}
\def\hybrid{\topmargin -20pt    \oddsidemargin 0pt
        \headheight 0pt \headsep 0pt
        \textwidth 6.25in       % A4 paper
        \textheight 9.25in       % A4 paper
        \marginparwidth .875in
        \parskip 5pt plus 1pt   \jot = 1.5ex}

\hybrid

\def\baselinestretch{1.2}

\catcode`\@=11

\def\marginnote#1{}
\newcount\hour
\newcount\minute
\newtoks\amorpm
\hour=\time\divide\hour by60
\minute=\time{\multiply\hour by60 \global\advance\minute by-\hour}
\edef\standardtime{{\ifnum\hour<12 \global\amorpm={am}%
        \else\global\amorpm={pm}\advance\hour by-12 \fi
        \ifnum\hour=0 \hour=12 \fi
        \number\hour:\ifnum\minute<10 0\fi\number\minute\the\amorpm}}
\edef\militarytime{\number\hour:\ifnum\minute<10 0\fi\number\minute}
\def\draftlabel#1{{\@bsphack\if@filesw {\let\thepage\relax
   \xdef\@gtempa{\write\@auxout{\string
      \newlabel{#1}{{\@currentlabel}{\thepage}}}}}\@gtempa
   \if@nobreak \ifvmode\nobreak\fi\fi\fi\@esphack}
        \gdef\@eqnlabel{#1}}
\def\@eqnlabel{}
\def\@vacuum{}
\def\draftmarginnote#1{\marginpar{\raggedright\scriptsize\tt#1}}

\def\draft{\oddsidemargin -.5truein
        \def\@oddfoot{\sl preliminary draft \hfil
        \rm\thepage\hfil\sl\today\quad\militarytime}
        \let\@evenfoot\@oddfoot \overfullrule 3pt
        \let\label=\draftlabel
        \let\marginnote=\draftmarginnote
   \def\@eqnnum{(\theequation)\rlap{\kern\marginparsep\tt\@eqnlabel}%
\global\let\@eqnlabel\@vacuum}  }

\def\preprint{\twocolumn\sloppy\flushbottom\parindent 2em
        \leftmargini 2em\leftmarginv .5em\leftmarginvi .5em
        \oddsidemargin -.5in    \evensidemargin -.5in
        \columnsep .4in \footheight 0pt
        \textwidth 10.in        \topmargin  -.4in
        \headheight 12pt \topskip .4in
        \textheight 6.9in \footskip 0pt
        \def\@oddhead{\thepage\hfil\addtocounter{page}{1}\thepage}
        \let\@evenhead\@oddhead \def\@oddfoot{} \def\@evenfoot{} }

\def\numberbysection{\@addtoreset{equation}{section}
        \def\theequation{\thesection.\arabic{equation}}}

\def\underline#1{\relax\ifmmode\@@underline#1\else
        $\@@underline{\hbox{#1}}$\relax\fi}

\def\titlepage{\@restonecolfalse\if@twocolumn\@restonecoltrue\onecolumn
     \else \newpage \fi \thispagestyle{empty}\c@page\z@
        \def\thefootnote{\fnsymbol{footnote}} }

\def\endtitlepage{\if@restonecol\twocolumn \else \newpage \fi
        \def\thefootnote{\arabic{footnote}}
        \setcounter{footnote}{0}}  %\c@footnote\z@ }

\catcode`@=12
\relax

\def\figcap{\section*{Figure Captions\markboth
        {FIGURECAPTIONS}{FIGURECAPTIONS}}\list
        {Figure \arabic{enumi}:\hfill}{\settowidth\labelwidth{Figure
999:}
        \leftmargin\labelwidth
        \advance\leftmargin\labelsep\usecounter{enumi}}}
 \relax
\def\tablecap{\section*{Table Captions\markboth
        {TABLECAPTIONS}{TABLECAPTIONS}}\list
        {Table \arabic{enumi}:\hfill}{\settowidth\labelwidth{Table
999:}
        \leftmargin\labelwidth
        \advance\leftmargin\labelsep\usecounter{enumi}}}
 \relax
\def\reflist{\section*{References\markboth
        {REFLIST}{REFLIST}}\list
        {[\arabic{enumi}]\hfill}{\settowidth\labelwidth{[999]}
        \leftmargin\labelwidth
        \advance\leftmargin\labelsep\usecounter{enumi}}}
 \relax
\makeatletter
\newcounter{pubctr}
\def\publist{\@ifnextchar[{\@publist}{\@@publist}}
\def\@publist[#1]{\list
        {[\arabic{pubctr}]\hfill}{\settowidth\labelwidth{[999]}
        \leftmargin\labelwidth
        \advance\leftmargin\labelsep
        \@nmbrlisttrue\def\@listctr{pubctr}
        \setcounter{pubctr}{#1}\addtocounter{pubctr}{-1}}}
\def\@@publist{\list
        {[\arabic{pubctr}]\hfill}{\settowidth\labelwidth{[999]}
        \leftmargin\labelwidth
        \advance\leftmargin\labelsep
        \@nmbrlisttrue\def\@listctr{pubctr}}}
 \relax
\makeatother
\newskip\humongous \humongous=0pt plus 1000pt minus 1000pt

\newif\ifdtup

\relax

\def\be{\begin{equation}}
\def\ee{\end{equation}}
\def\ba{\begin{eqnarray}}
\def\ea{\end{eqnarray}}

\def\a{\alpha}

\def\b{\beta}

\def\g{\gamma}
\def\G{\Gamma}
\def\d{\delta}
\def\D{\Delta}

\def\m{\mu}

\def\l{\lambda}
\def\L{\Lambda}
\def\s{\sigma}

\def\cN{{\cal N}}

\def\no{\noindent}

\def\qq{\qquad}

\def\IR{\relax{\rm I\kern-.18em R}}

\def \ov {\over}

\def\IR{\relax{\rm I\kern-.18em R}}
\def\IL{\relax{\rm I\kern-.18em L}}

\def\inv{^{\raise.15ex\hbox{${\scriptscriptstyle -}$}\kern-.05em 1}}

\def\Tr{{\rm Tr}}

\begin{document}
%Text\fontsize{13}{12}\selectfont Text

\allowdisplaybreaks
\renewcommand{\theequation}{\thesection.\arabic{equation}}
\csname @addtoreset\endcsname{equation}{section}

\newcommand{\beq}{\begin{equation}}
\newcommand{\eeq}[1]{\label{#1}\end{equation}}
\newcommand{\ber}{\begin{equation}}
\newcommand{\eer}[1]{\label{#1}\end{equation}}
\newcommand{\eqn}[1]{(\ref{#1})}
\begin{titlepage}
\begin{center}

%\hfill CERN-TH-2017-148
%\vskip -.1 cm
%\hfill hep--th/yymmnnn\\
%\hfill JHEP {\bf 1711} (2017) 078

%\today
${}$
\vskip .2 in

\vskip .4cm

{\large\bf
A generalized method for all-loop results
 in $\l$-models
}

\vskip 0.4in

 {\bf Eftychia Sagkrioti}
\vskip 0.16in

 {\em
Department of Nuclear and Particle Physics,\\
Faculty of Physics, National and Kapodistrian University of Athens,\\
Athens 15784, Greece\\
}

\vskip 0.12in

{\footnotesize \texttt {esagkrioti@phys.uoa.gr,\\ felicity.sagkriotis@gmail.com}}

%\today

\vskip .5in
\end{center}

\centerline{\bf Abstract}

\no
We compute the anomalous dimension of the single current operator in the case of single and doubly deformed asymmetric $\l$-models with a general deformation matrix.
Our method uses the underlying geometry of the coupling space, as well as an auxiliary group interaction, which completely decouples from the asymmetric model in a specific limit, consistent with the 
Renormalization Group flow. 
Our results are valid to all orders in the deformation parameters and leading order to the levels of the underlying current algebras. 
We specialize our general result to several models of particular interest that have been
constructed in the literature and for which these anomalous dimensions were not known.

\vskip .4in
\noindent
\end{titlepage}
\vfill
\eject

\newpage
%\vskip .3in

\tableofcontents

\noindent

\def\baselinestretch{1.2}
\baselineskip 20 pt
\noindent

%%%%%%%%%%%%%%%

\setcounter{equation}{0}

\section{Introduction}
A few years ago, a new class of CFT deformations has been introduced in \cite{Sfetsos:2013wia}, describing interpolations between exact CFTs and non-Abelian T-duals of Principal Chiral Models or geometric coset models.  For a semi-simple, compact Lie group G, the so called $\l$-models are naturally constructed by gauging symmetries of
integrable models \cite{Sfetsos:2013wia}, ending up with an effective action which includes all $\l$-dependent quantum corrections
\begin{equation}
S_{\l}(g)=S_{k}(g)+\frac{k}{\pi}\int d^2\s J_+^a(\l^{-1}-D^T)^{-1}_{ab}J_-^b, \label{l-model}
\end{equation}
where $S_k$ is the WZW action at level k
\begin{equation}
S_k(g)=\frac{k}{2\pi}\int\Tr(g^{-1}\partial_+gg^{-1}\partial_-g)+\frac{k}{12\pi}\int_B\Tr(g^{-1}\text{d}g)^3\,,
\end{equation}
and
\begin{align}
\begin{split}
J^a_+=-i\Tr(t^a\partial_+gg^{-1}),\qq
J^a_-=-i\Tr(t^ag^{-1}\partial_-g),\qq
D_{ab}=\Tr(t_agt^bg^{-1}).
\end{split}
\end{align}
The matrices $t^a$ are hermitian generators of $\mathfrak{g}=\text{Lie}(G)$ satisfying the commutation relations $[t^a,t^b]=if_{abc}t^c$ with real structure constants and are normalized as $\Tr(t^at^b)=\d^{ab}$. Thus, in the adjoint representation the following relation holds $f_{acd}f_{bcd}=c_G\d_{ab}$, where $c_G$ is the eigenvalue of the quadratic Casimir operator in this representation. \\
The effective theory \eqref{l-model} has an additional remarkable duality-type symmetry, involving the coupling matrix $\l_{ab}$, the level k of the Kac-Moody algebra $\mathfrak{g}_k$ and the group element $g\in G$
\begin{equation}
\l\to\l^{-1},\qq k\to -k,\qq g\to g^{-1}, \label{symmetry}
\end{equation}
which is also reflected to physical quantities.
For small values of $\l_{ab}$, the linearised form of the action \eqref{l-model} is the (generalized) non-Abelian Thirring model
\begin{equation}
S_{\l}(g)=S_{k}(g)+\frac{k}{\pi}\int d^2\s \l_{ab}J^a_+J^b_-+\mathcal{O}(\l^2), \label{perturbed l-model}
\end{equation}
corresponding to a WZW model at level k, perturbed by a set of classically marginal, current bilinear operators. We should note here, that the level k is of topological nature and does not run under the Renormalization Group. The correlation functions and RG flows for the model \eqref{perturbed l-model}, with a diagonal and isotropic version of deformation matrix, have been extensively analysed in \cite{Georgiou:2016iom} and \cite{Georgiou:2015nka}, while the case for general coupling matrix and symmetric coset space are presented in \cite{Itsios:2014lca,Sfetsos:2014jfa}. \\
Further generalizations of $\l$-deformed models followed after the original construction, including the left-right asymmetric case \cite{Georgiou:2016zyo} of the initial model \cite{Sfetsos:2013wia}, as well as symmetric \cite{Georgiou:2016urf,Georgiou:2017aei}  and asymmetric, double  or cyclic deformations \cite{Georgiou:2017oly,Sagkrioti:2018rwg} involving more interacting WZW models with different current algebras. \\
\\
In what follows we consider the case of the doubly deformed asymmetric model with a general deformation matrix \cite{Sagkrioti:2018rwg}, in order to present a new method to obtain the anomalous dimension of the single currents without using perturbation theory. For a diagonal deformation matrix, the usual method to compute the fundamental current anomalous dimension is with the use of perturbation theory combined with the non-perturbative duality-type symmetry \cite{Georgiou:2015nka,Georgiou:2016zyo,Georgiou:2017aei}. However, for the case of a general deformation matrix, the aforementioned method cannot be applied and no other approach is known in order to obtain an exact result for the single current anomalous dimension\footnote{One could proceed by purely using the method developed in Sections 2,3 and 4 of \cite{exact results}. However, this procedure would not take into account the contribution arising from diffeomorphisms of the target space, which in many cases prove to be crucial in order to obtain a consistent theory, given a reduced form for the matrix $\l_{a\hat{b}}$.}.  The method presented in this work provides a solution to this problem by only taking advantage of the couplings' space geometrical data and is based on a combination of the two methods analysed in \cite{exact results}. As we will show, the results of this method can be naturally modified in order to include the effects of target space diffeomorphisms, when the latter are needed. \\
The plan of this paper is the following:
In Section 2 the essential features of our method are introduced. In Section 3 we compute the anomalous dimensions of the composite bilinear operators and employ our method to derive the anomalous dimensions of the fundamental currents. In Section 4 we consider the two couplings case, using a subgroup and a general coset, and compute the anomalous dimensions for the
% coset and subgroup
corresponding single currents. In Section 5 the $SU(2)$ example with diagonal and fully anisotropic deformation matrix is presented, along with the case of a more general, non-diagonal $\l$ matrix.
Finally, the case of anomalous dimensions with diffeomorphisms is analysed in Appendix A.
\section{Setting up the frame}
Our starting point is the doubly deformed asymmetric case involving two copies of the semi-simple Lie group $G$, with different Kac-Moody algebra levels $k_1$, $k_2$, such that the group elements $g_1\in G_{k_1}$, $g_2\in G_{k_2}$. The linearised action of this model is \cite{Sagkrioti:2018rwg}
\begin{equation}
S_{\l_1,\l_2}=S_{k_1}(g_1)+S_{k_2}(g_2)+\frac{\sqrt{k_1k_2}}{\pi}\int d^2 \s \Big((\l_1)_{a\hat{b}}J_{1+}^aJ_{2-}^{\hat{b}}+(\l_2)_{\hat{a}b}J_{2+}^{\hat{a}}J_{1-}^b\Big)+\dots\,, \label{action1}
\end{equation}
where the hatted indices denote elements of the second copy of the group.\\
Its all-loop effective form can be found in \cite{Sagkrioti:2018rwg,Georgiou:2017jfi} and is invariant under the generalized duality-type symmetry
\begin{equation}
k_1\to-k_2,\quad k_1\to-k_1,\quad \l_1\to \l_1^{-1},\quad \l_2\to \l_2^{-1},\quad g_1\to g_2^{-1},\quad g_2\to g_1^{-1}\  \label{generalised symmetry}
\end{equation}
and the generalized parity transformation
\begin{equation}
\s\to-\s, \quad \l_i\to\l_i^T,\quad k_1\to k_2,\quad k_2\to k_2,\quad g_1\to g_2^{-1},\quad g_2\to g_1^{-1}\ . \label{parity}
\end{equation}
In the $\l_2=0$ limit\footnote{As it has been proved in
	% \cite{Georgiou:2016zyo}, \cite{Georgiou:2017aei} and \cite{c-function:2018},
 \cite{Georgiou:2017aei,Georgiou:2017oly,Sagkrioti:2018rwg}%,\cite{Georgiou:2017oly}, \cite{Sagkrioti:2018rwg}
  and  \cite{Georgiou:2017jfi} the two flows of (\ref{action1}) are decoupled and thus this limit is valid, with the results for \eqref{action1} corresponding to two copies of the \eqref{action2} ones.} and by renaming $\l_1=\l$, \eqref{action1} becomes
\begin{equation}
	S_{\l}(g_1,g_2)=S_{k_1}(g_1)+S_{k_2}(g_2)+\frac{\sqrt{k_1k_2}}{\pi}\int d^2 \s \l_{a\hat{b}}J_{1+}^aJ_{2-}^{\hat{b}}\ , \label{action2}
\end{equation}
which is now exact in the parameter $\l$ \cite{Georgiou:2017jfi}.\\
Our goal is to compute the anomalous dimension of $J_{1+}$ and $J_{2-}$ currents for the model described by \eqref{action2}.
To do so, we add a new interaction term with coupling $\tilde{\l}_{a\tilde{b}}$, involving a third copy $G_{k_3}$ of the group $G$, with Kac-Moody currents $J_{3\pm}$ and consider the following action
\begin{equation}
S_{\l,\tilde{\l}}(g_1,g_2,g_3)=\sum_{i=1}^{3}S_{k_i}(g_i)+\frac{\sqrt{k_1k_2}}{\pi}\int d^2 \s \l_{a\hat{b}}J_{1+}^aJ_{2-}^{\hat{b}}+\frac{\sqrt{k_1k_3}}{\pi}\int d^2 \s \tilde{\l}_{a\tilde{b}}J^a_{1+}J^{\tilde{b}}_{3-}\,, \label{action3}
\end{equation}
where now the tilde indices label the elements of the third group copy.
Notice here that \eqref{action3} again corresponds to an effective action, incorporating all quantum corrections both for $\l$ and $\tilde{\l}$ \cite{Georgiou:2018gpe}.
\begin{comment}
Since the three copies of the group $G$ are considered to be independent from each other, the different chiral currents, after the rescaling $J_{i+}^a\to J_{i+}^a/\sqrt{k_i}$, obey the OPE
\begin{align}
\begin{split}
J_{i+}^a(z)J_j^b(w)=\d_{ij}\left(\frac{\d^{ab}}{(z-w)^2}+\frac{f^{abc}}{\sqrt{k_j}}\frac{J_j^c(w)}{z-w}\right),\qq i,j=1,2,3 \label{OPEs}
\end{split}
\end{align}
and an equivalent relation holds for the anti-chiral ones as well, while $J_{i+}^a(z)J_{j-}^b(\bar{w})=\text{regular}$.
\end{comment}
\\
\\
We can now compute the anomalous dimension of the two composite operators, along with the anomalous dimension of the single current operator $J_{1+}$.\\
The idea is the following: We can bring the action (\ref{action3}) to the form \cite{Georgiou:2018gpe}
\begin{equation}
S_{\l,\tilde{\l}}(g_1,g_2,g_3)=\sum_{i=1}^{3}S_{k_i}(g_i)+\frac{1}{\pi}\int d^2\s\mathcal{J}_+^A\L_{AB}\mathcal{J}_-^B, \label{action new form}
\end{equation}
where the currents have been rescaled as $J_{i\pm}^a\to J_{i\pm}^a/\sqrt{k_i}$ and $\mathcal{J}_{\pm}^A=(J_{1\pm}^a,J_{2\pm}^{\hat{a}},J_{3\pm}^{\tilde{a}})$. Here, the triple index notation $A=(a,\hat{a},\tilde{a})$ has been used to denote the indices of the first, second and third copy of $G$ respectively, and the matrix $\L$ now is\footnote{The matrix $\L$ here is not invertible. However this doesn't affect our results, since no inversion is needed in the present context.}
\begin{equation}
\label{L1}
	\L_{AB}=\begin{pmatrix}0 && \l_{a\hat{b}} && \tilde{\l}_{a\tilde{b}}\\
	0 && 0 && 0\\
	0 && 0 && 0
\end{pmatrix}.
\end{equation}
For this model we can now compute the anomalous dimension of the two composite operators $J_{1+}J_{2-}$ and $J_{1+}J_{3-}$ by using the couplings' space geometry. Then, by taking the limit $\tilde{\l}_{ab}=0$, $J_{3-}$ decouples from the action and due to having a trivial OPE with the other operators in it, it does not acquire an anomalous dimension. Thus, we can find the anomalous dimension of the single current $J_{1+}$ from the anomalous dimension of the $J_{1+}J_{3-}$ composite operator. This can be done consistently, since in the aforementioned limit which is also consistent with the RG flows, the anomalous dimension matrix for the bilinear operators proves to be block-diagonal, implying no mixing between composite operators belonging in different blocks.
% Thus, in the decoupling limit, all results containing $G_{k_3}$ indices with tildes, are expected to be proportional to $\d_{\tilde{a}}{}^{\tilde{b}}$.
\\
\\
Due to \eqref{parity}, it is clear that one can obtain the anomalous dimension of $J_{2-}$ from the one of $J_{1+}$, by replacing $\l$ with $\l^T$ and exchanging $k_1$ with $k_2$.
\\
However, a more strict way to compute the anomalous dimension of $J_{2-}$, would be to modify the previous procedure accordingly.
%In order to compute the anomalous dimension for the current $J_{2-}$ of the second group, a small modification is needed.
 In this case, the auxiliary interaction term added to (\ref{action2}) has to be of the form $\frac{\sqrt{k_1k_3}}{\pi}\tilde{\l}_{\tilde{a}\hat{b}}J^{\tilde{a}}_{3+}J^{\hat{b}}_{2-}$, such that the anomalous dimension of $J_{3+}^{\tilde{a}}J_{2-}^{\hat{b}}$ reduces to the one of $J_{2-}$ in the $\tilde{\l}_{a\tilde{b}}=0$ limit. Therefore, the $\L$ matrix will be of the following form
\begin{equation}
\label{L2}
\L_{AB}=\begin{pmatrix}0 && \l_{a\hat{b}} && 0\\
0 && 0 && 0\\
0 && \tilde{\l}_{\tilde{a}\hat{b}} && 0
\end{pmatrix}\ .
\end{equation}
Finally, using the results for the anomalous dimensions of ${J_{1+}}$ and ${J_{2-}}$, it is easy to obtain the anomalous dimensions for the single currents entering in the second interaction vertex of \eqref{action1}, just by replacing the matrix $\l_1$ with $\l_2$ and $k_1\leftrightarrow k_2$. \\In the symmetric limit $k_1=k_2$, all the aforementioned results reduce to the corresponding ones for the simply deformed $\l$-model of \eqref{l-model} and \eqref{perturbed l-model}.

\section{Anomalous dimensions}
In this Section we compute the anomalous dimension for the $J_{1+}$ %and $J_{2-}$ currents 
current and the final result is presented in \eqref{andimJ1J3}.  The general formula for the anomalous dimensions of the composite operators that drive the theory away from the CFT point, can be found from the two point function \cite{Georgiou:2015nka,Kutasov:1989dt}
\begin{equation}
\langle{J^A_{1+}J^B_{2-}(x_1,\bar x_1)J^C_{1+}J^D_{2-}(x_2,\bar x_2)}\rangle_{\l,k}
=\frac{G_{CD|MN}}{|x_{12}|^4}\left(\d_A{}^M\d_B{}^N+\g_{AB}{}^{MN}\ln\frac{\varepsilon^2}{|x_{12}|^2}\right)\ ,
\end{equation}
and reads \cite{Sagkrioti:2018abh}
\begin{equation}
\g_{AB}{}^{CD}=\nabla_{AB}\b^{CD}+\nabla^{CD}\b_{AB}=\nabla_{AB}\b^{CD}+G_{AB|MN}G^{CD|PQ}\nabla_{PQ}\b^{MN}\ ,\label{andimgeneral}
\end{equation}
where
\begin{equation}
 \nabla_{AB}\b^{CD}=\partial_{AB}\b^{CD}+\G_{AB|MN}^{CD}\b^{MN},\qq \partial_{AB}\b^{CD}=\frac{\partial\b^{CD}}{\partial\L_{AB}},
 \end{equation}
is the covariant derivative defined in the space of couplings of the matrix $\L$\footnote{In order to follow the bibliography, we define the coordinate space elements $\L_{AB}$ with both indices down.}. The Christoffel symbols used in the above expression are the usual ones and are defined with respect to the Zamolodchikov's metric $G_{AB|CD}$ of the couplings' space \cite{Sagkrioti:2018abh}
\begin{align}
\begin{split}
&G_{AB|CD}=\frac{1}{2}\tilde{g}^{AC}g^{BD},\quad \tilde{g}_{AB}=(1-\L\L^T)_{AB},\quad g_{AB}=(1-\L^T\L)_{AB},\\
&\tilde{g}^{AB}=\tilde{g}_{AB}^{-1},\qq g^{AB}=g_{AB}^{-1},\qq G_{AB|MN}G^{MN|CD}=\d_A{}^C\d_B{}^D\,. \label{Zamolo metric definition}
\end{split}
\end{align}
\begin{comment}
and read
\begin{align*}
\begin{split}
\G_{M_1M_2|N_1N_2}^{P_1P_2}=\frac{1}{2}G^{P_1P_2|Q_1Q_2}\left(\partial_{M_1M_2}G_{Q_1Q_2|N_1N_2}
+\partial_{N_1N_2}G_{Q_1Q_2|M_1M_2}-\partial_{Q_1Q_2}G_{M_1M_2|N_1N_2}\right)\,.
\end{split}
\end{align*}
\end{comment}
By substituting \eqref{Zamolo metric definition} and the covariant derivatives inside \eqref{andimgeneral}, we get the following simplified formula for the anomalous dimension of the composite operators
\begin{equation}
	\g_{AB}{}^{CD}=\partial_{AB}\b^{CD}+\tilde{g}_{CP}g_{DQ}\bigg(\tilde{g}^{AM}g^{BN}\partial_{PQ}\b^{MN}+\b^{KL}\partial_{KL}\big(\tilde{g}^{PA}g^{QB}\big)\bigg)\,. \label{andimgeneralwithGammasSubstituted}
\end{equation}
Also, for the $\b$-functions we have
\begin{align}
\begin{split}
\b^{AB}=\frac{\text{d}\L_{AB}}{\text{d}\ln\m^2}, \qq \b_{AB}=G_{AB|CD}\b^{CD}\,,
\end{split}
\end{align}
where $\m$ is the energy scale.
By taking the levels $k_1=k_2=k_3=1$ and then restoring them through a redefinition of the structure constants \cite{Georgiou:2018gpe} of the three copies of the algebra as $F_{ABC}=\left(f_{abc}/\sqrt{k_1}, f_{\hat{a}\hat{b}\hat{c}}/\sqrt{k_2},f_{\tilde{a}\tilde{b}\tilde{c}}/\sqrt{k_3}  \right)$\footnote{All other components of $F_{ABC}$ with mixed indices are zero since the three copies of the group are considered independent. Moreover, the structure constants $F_{ABC}$ are real.},
it is easy to bring the $\b$-functions of \cite{Sagkrioti:2018rwg} in the following form
\begin{align}
\begin{split}
&\b^{AB}=\frac{1}{2}\mathcal{N}_{AC}{}^D(\L)\mathcal{N}_{BD}{}^C(\L^T),\quad \mathcal{N}_{AB}{}^C(\L)=(\L_{AE}\L_{BD}F_{EDF}-\L_{EF}F_{ABE})g^{FC}. \label{betawithNs}
\end{split}
\end{align}

\subsection{Anomalous dimension of the single current}
In what follows, we firstly compute the exact in $\l$ and $\tilde{\l}$, and up to order $\mathcal{O}(1/k)$, $\b$-functions for the two blocks of couplings appearing in \eqref{action3}. Subsequently, the exact in $\l$ and up to $\mathcal{O}(\tilde{\l}^0)$ anomalous dimensions of the composite operators $J_{1+}J_{2-}$, $J_{1+}J_{3-}$ are found, from which the anomalous dimension of the single currents $J_{1+}$ %, $J_{2-}$ 
follows in the decoupling limit.
\subsubsection{The exact in all couplings $\b$-functions}
For the case of \eqref{action3} the $\L_{AB}$ matrix is the one in (\ref{L1}) and the $\b$-function will be of the form
\begin{equation}
\b^{AB}=\frac{\text{d}\L_{AB}}{\text{d}\ln\m^2}=\begin{pmatrix}
0 && \b^{a\hat{b}} && \b^{a\tilde{b}}\\
0 && 0 && 0\\
0 && 0 && 0
\end{pmatrix}, \label{matrix beta 1}
\end{equation}
where
\begin{equation}
\b^{a\hat{b}}=\frac{1}{2}\mathcal{N}_{aC}{}^D(\L)\mathcal{N}_{\hat{b}D}{}^C(\L^T),\quad\b^{a\tilde{b}}=\frac{1}{2}\mathcal{N}_{aC}{}^D(\L)\mathcal{N}_{\tilde{b}D}{}^C(\L^T), \label{bl and blt with Ns}
\end{equation}
and
\small
	\begin{align}
&\L_{AB}= \begin{pmatrix}
0 && \l_{a\hat{b}} && \tilde{\l}_{a\tilde{b}} \\
0 && 0 && 0\\
0 && 0 && 0
\end{pmatrix},\qq
(\L^T)_{AB}=\begin{pmatrix}
0 && 0 && 0 \\
(\l^T)_{\hat{a}b} && 0 && 0\\
(\tilde{\l}^T)_{\tilde{a}b} && 0 && 0
\end{pmatrix},
 \label{l,g,gt,d,f,h}\\
& g_{AB}=\begin{pmatrix}
\d_{ab} && 0 && 0 \\
0 && (1-\l^T\l)_{\hat{a}\hat{b}} && -(\l^T\tilde{\l})_{\hat{a}\tilde{b}}\\
0 && -(\tilde{\l}^T\l)_{\tilde{a}\hat{b}} && (1-\tilde{\l}^T\tilde{\l})_{\tilde{a}\tilde{b}}
\end{pmatrix},\quad
 g^{AB}=\begin{pmatrix}
\d_{ab} && 0 && 0 \\
0 && f(\l,\tilde{\l})_{\hat{a}\hat{b}} && h(\l,\tilde{\l})_{\hat{a}\tilde{b}}\\
0 && h(\tilde{\l},\l)_{\tilde{a}\hat{b}} && f(\tilde{\l},\l)_{\tilde{a}\tilde{b}}
\end{pmatrix},\nonumber\\
& \tilde{g}_{AB}=\begin{pmatrix}
\D^{-1}_{ab} && 0 && 0 \\
0 && \d_{\hat{a}\hat{b}} && 0\\
0 && 0 && \d_{\tilde{a}\tilde{b}}
\end{pmatrix},\quad
\tilde{g}^{AB}=\begin{pmatrix}
\D_{ab} && 0 && 0 \\
0 && \d_{\hat{a}\hat{b}} && 0\\
0 && 0 && \d_{\tilde{a}\tilde{b}}
\end{pmatrix},\quad \D=(1-\l\l^T-\tilde{\l}\tilde{\l}^T)^{-1}, \nonumber\\
&f(\l,\tilde{\l})=\l^T\D(1-\tilde{\l}\tilde{\l}^T)\l^{-T},\qq h(\l,\tilde{\l})=\l^T\D\tilde{\l},\qq \l^{-T}=(\l^{-1})^T\ . \nonumber
\end{align}
\normalsize
The other components of (\ref{matrix beta 1}) are indeed zero as can also be confirmed from (\ref{betawithNs}) and thus no diffeomorphisms are needed to be added as counterterms, since no new directions are generated in the RG flow. However, inside the $\b^{a\hat{b}}$ and $\b^{a\tilde{b}}$ blocks, new flows may occur in general, depending on the specific choice of $\l$ and $\tilde{\l}$ matrices. These should be cancelled with the use of appropriate diffeomorphisms and this issue is addressed in Appendix A.\\
\\
Returning to \eqref{bl and blt with Ns} the non-zero components of $\mathcal{N}_{aB}{}^C(\L)$ and $\mathcal{N}_{\hat{a}B}{}^C(\L^T)$ are
\begin{align}
\begin{split}
&\mathcal{N}_{ab}{}^
{\hat{c}}(\L)=\frac{1}{\sqrt{k_2}}N_{ab}{}^{\hat{c}}(\l,\tilde{\l},\l_0^{-1})+\frac{1}{\sqrt{k_3}}\tilde{N}_{ab}{}^{\hat{c}}(\l,\tilde{\l},\tilde{\l}_0^{-1}),\\
&\mathcal{N}_{ab}{}^{\tilde{c}}(\L)=\frac{1}{\sqrt{k_2}}\tilde{N}_{ab}{}^{\tilde{c}}(\tilde{\l},\l,\l_0^{-1})+\frac{1}{\sqrt{k_3}}N_{ab}{}^{\tilde{c}}(\tilde{\l},\l,\tilde{\l}_0^{-1}),\\
&\mathcal{N}_{\hat{a}\hat{b}}{}^c(\L^T)=\frac{1}{\sqrt{k_1}}\mathcal{\tilde{N}}_{\hat{a}\hat{b}}{}^c(\l^T,\tilde{\l},\l_0),\\
&\mathcal{N}_{\hat{a}\tilde{b}}{}^c(\L^T)=\frac{1}{\sqrt{k_1}}(\l^T)_{\hat{a}e}(\tilde{\l}^T)_{\tilde{b}d}f_{edf}\D_{fc}\ , \label{Ns}
\end{split}
\end{align}
where
\begin{align}
\begin{split}
&N_{ab}{}^{\g}(A,B,\a)=(A_{ae}A_{bd}f_{edf}-\a A_{ef}f_{abe})(A^T\D(1-BB^T)A^{-T})_{f\g},\\
&\tilde{N}_{ab}{}^{\g}(A,B,\a)=(B_{ae}B_{bd}f_{edf}-\a B_{ef} f_{abe})(B^T\D A)_{f\g},\quad \text{with}\quad \g=(\hat{c},\tilde{c}),\\
&\mathcal{\tilde{N}}_{\hat{a}\hat{b}}{}^c(A,B,\a)=(A_{ae}A_{bd}f_{edf}-\a A_{ef}f_{abe})\D_{fc},\quad \D=\D(A,B),\\
&\l_0=\sqrt{\frac{k_1}{k_2}},\qq \tilde{\l}_0=\sqrt{\frac{k_1}{k_3}}. \label{myNs}
\end{split}
\end{align}
The exact in $\l$ and $\tilde{\l}$ $\b$-functions, up to $\mathcal{O}(1/k)$, are then
\small
\begin{align}
\begin{split}
\b^{a\hat{b}}&=\frac{1}{2\sqrt{k_1k_2}}\Bigg(N_{ac}{}^{\hat{d}}(\l,\tilde{\l},\l_0^{-1})\mathcal{\tilde{N}}_{\hat{b}\hat{d}}{}^c(\l^T,\tilde{\l},\l_0)+\tilde{N}_{ac}{}^{\tilde{d}}(\tilde{\l},\l,\l_0^{-1})\l^T_{\hat{b}e}\tilde{\l}^T_{\tilde{d}i}\D_{fc}f_{eif} \Bigg)\\
&+\frac{1}{2\sqrt{k_1k_3}}\Bigg(\tilde{N}_{ac}{}^{\hat{d}}(\l,\tilde{\l},\tilde{\l}_0^{-1})\mathcal{\tilde{N}}_{\hat{b}\hat{d}}{}^c(\l^T,\tilde{\l},\l_0)+N_{ac}{}^{\tilde{d}}(\tilde{\l},\l,\tilde{\l}_0^{-1})\l^T_{\hat{b}e}\tilde{\l}^T_{\tilde{d}i}\D_{fc}f_{eif}  \Bigg)\ ,\label{bl}
\end{split}
\end{align}
\begin{align}
\begin{split}
\b^{a\tilde{b}}&=\frac{1}{2\sqrt{k_1k_2}}\Bigg(N_{ac}{}^{\hat{d}}(\l,\tilde{\l},\l_0^{-1})\tilde{\l}^T_{\tilde{b}e}\l^T_{\hat{d}i}\D_{fc}f_{eif} +\tilde{N}_{ac}{}^{\tilde{d}}(\tilde{\l},\l,\l_0^{-1})\tilde{\mathcal{N}}_{\tilde{b}\tilde{d}}{}^c(\tilde{\l}^T,\l,\tilde{\l}_0) \Bigg)\\
	&+\frac{1}{2\sqrt{k_1k_3}}\Bigg(\tilde{N}_{ac}{}^{\hat{d}}(\l,\tilde{\l},\tilde{\l}_0^{-1})\tilde{\l}^T_{\tilde{b}e}\l^T_{\hat{d}i}\D_{fc}f_{eif}+N_{ac}{}^{\tilde{d}}(\tilde{\l},\l,\tilde{\l}_0^{-1})\tilde{\mathcal{N}}_{\tilde{b}\tilde{d}}{}^c(\tilde{\l}^T,\l,\tilde{\l}_0)  \Bigg)\ .\label{blt}
\end{split}
\end{align}
\normalsize
We may now consider the above $\b$-functions at some important limits:\footnote{We will use here that $f_{acd}f_{bcd}=c_G\d_{ab}$.%, with $c_G$ being the eigenvalue of the quadratic Casimir operator in the adjoint representation.
}
\begin{itemize}
	\item In the diagonal $\l_{a\hat{b}}=\l\d_{ab}$ and $\tilde{\l}_{a\tilde{b}}=\tilde{\l}\d_{ab}$ limit, the exact in $\l$ and $\tilde{\l}$ $\b$-functions read
	\begin{align}
	%\begin{split}
	&\b_{\l}(\l,\tilde{\l})=-\frac{c_G}{2}\Bigg(\frac{f_1(\l,\tilde{\l},\l_0)}{\sqrt{k_1k_2}}+\frac{f_2(\l,\tilde{\l},\tilde{\l}_0)}{\sqrt{k_1k_3}}\Bigg),\nonumber\\ &\b_{\tilde{\l}}(\l,\tilde{\l})=-\frac{c_G}{2}\Bigg(\frac{f_2(\tilde{\l},\l,\l_0)}{\sqrt{k_1k_2}}+\frac{f_1(\tilde{\l},\l,\tilde{\l}_0)}{\sqrt{k_1k_3}}\Bigg), \label{l, lt diagonal limit}\\
	&f_1(\l,\tilde{\l},\l_0)=\l^2(\l-\l_0^{-1})(\l-\l_0+\l_0\tilde{\l}^2)\D^2_{\text{diag}},\nonumber\\
	&f_2(\l_,\tilde{\l},\tilde{\l}_0)=\l\tilde{\l}^2(\tilde{\l}-\tilde{\l}_0^{-1})(1-\l_0\l)\D^2_{\text{diag}},\qq \D_{\text{diag}}=(1-\l^2-\tilde{\l}^2)^{-1}.\nonumber
%	\end{split}
	\end{align}
	Notice here that the $\b$-functions of \eqref{bl} and \eqref{blt} are coupled even in the diagonal limit. This is expected despite the fact that the OPEs between currents belonging to different copies of the group $G$ vanish, since both interaction vertices of \eqref{action3} contain the chiral current $J_{1+}$ which has non-zero OPE with itself. Furthermore, \eqref{l, lt diagonal limit} coincides with eq. (4.7) of \cite{Georgiou:2018gpe} for $n=4$ and the redefinition $\l\to \l_0\l$ and $\tilde{\l}\to \tilde{\l}_0\tilde{\l}$ of the couplings, as expected.
	%This is expected since from the OPEs \eqref{OPEs} becomes clear that the currents of two interaction vertices of \eqref{action3} can be contracted.
	\item In the $\tilde{\l}=0$ and $\l_{a\hat{b}}=\l\d_{ab}$ limit
		\begin{align}
	\begin{split}
	&\b_{\l}(\l,0,\l_0)=-\frac{c_G}{2\sqrt{k_1k_2}}\frac{\l^2(\l-\l_0^{-1})(\l-\l_0)}{(1-\l^2)^2},\qq \b_{\tilde{\l}}(\l,0,\tilde{\l}_0)=0, \label{beta doubly deformed asymmetric diagonal}
	\end{split}
	\end{align}
	with the $\b_{\l}$ corresponding to the $\b$-function of the doubly deformed model \cite{Georgiou:2016zyo,Sagkrioti:2018rwg} with group copies $G_{k_1}$ and $G_{k_2}$ as expected.
	\item  In the $\l=0$ and $\tilde{\l}_{a\tilde{b}}=\tilde{\l}\d_{ab}$ limit, $\b_\l(0,\tilde{\l},\l_0)=0$, while $\b_{\tilde{\l}}(0,\tilde{\l},\tilde{\l}_0)$ is the same as the $\b_{\l}$ of \eqref{beta doubly deformed asymmetric diagonal} but with $k_2$ replaced by $k_3$ and $\l$ by $\tilde{\l}$,
\begin{comment}
	\begin{equation}
	\b_\l(0,\tilde{\l},\l_0)=0, \qq \b_{\tilde{\l}}(0,\tilde{\l},\tilde{\l}_0)=-\frac{c_G}{2\sqrt{k_1k_3}}\frac{\tilde{\l}^2(\tilde{\l}-\tilde{\l}_0^{-1})(\tilde{\l}-\tilde{\l}_0)}{(1-\tilde{\l}^2)^2}, \label{l=0 diagonal lt}
	\end{equation}
\end{comment}
	corresponding again to the doubly deformed, asymmetric model, but now with group copies $G_{k_1}$ and $G_{k_3}$, as expected.
\end{itemize}

\subsubsection{Anomalous dimension in the decoupling limit}
Since we are interested in computing the anomalous dimension of $J_{1+}$ for the $\l$-deformed model of \eqref{action2}, we can work directly in the $\tilde{\l}=0$ limit. In this limit, only $\b^{a\hat{b}}$ is non-zero, and \eqref{andimgeneralwithGammasSubstituted} becomes
\begin{equation}
\g_{AB}{}^{CD}=\partial_{AB}\b^{CD}+\tilde{g}_{CP}g_{DQ}\bigg(\tilde{g}^{AM}g^{BN}\partial_{PQ}\b^{MN}+\b^{k\hat{l}}\partial_{k\hat{l}}\big(\tilde{g}^{PA}g^{QB}\big)\bigg)\,. \label{andimgeneralwithGammasSubstitutedintheDecouplingLimit}
\end{equation}
Due to the possible appearance of derivatives with respect to $\tilde{\l}$ acting on the $\b$-functions, only $\mathcal{O}(\tilde{\l})$ terms in the $\b$-functions and $\mathcal{O}(\tilde{\l}^0)$ in the Zamolodchikov's metric will contribute for $\tilde{\l}=0$.
Thus, in this limit, \eqref{l,g,gt,d,f,h} reduces to
\begin{align}
&\L_{AB}= \begin{pmatrix}
0 && \l_{a\hat{b}} && 0 \\
0 && 0 && 0\\
0 && 0 && 0
\end{pmatrix},\qq
(\L^T)_{AB}=\begin{pmatrix}
0 && 0 && 0 \\
(\l^T)_{\hat{a}b} && 0 && 0\\
0 && 0 && 0
\end{pmatrix}, \nonumber\\
& g_{AB}=\begin{pmatrix}
\d_{ab} && 0 && 0 \\
0 && (1-\l^T\l)_{\hat{a}\hat{b}} && 0\\
0 && 0 && \d_{\tilde{a}\tilde{b}}
\end{pmatrix},\quad
g^{AB}=\begin{pmatrix}
\d_{ab} && 0 && 0 \\
0 && (1-\l^T\l)^{-1}_{\hat{a}\hat{b}} && 0\\
0 && 0 && \d_{\tilde{a}\tilde{b}}
\end{pmatrix}, \label{1st order}\\
& \tilde{g}_{AB}=\begin{pmatrix}
\D^{-1}_{ab} && 0 && 0 \\
0 && \d_{\hat{a}\hat{b}} && 0\\
0 && 0 && \d_{\tilde{a}\tilde{b}}
\end{pmatrix},\quad
\tilde{g}^{AB}=\begin{pmatrix}
\D_{ab} && 0 && 0 \\
0 && \d_{\hat{a}\hat{b}} && 0\\
0 && 0 && \d_{\tilde{a}\tilde{b}}
\end{pmatrix},\quad \D=(1-\l\l^T)^{-1} . \nonumber
\end{align}
Then, the $\b$-functions \eqref{bl}, \eqref{blt} up to $\mathcal{O}(\tilde{\l})$ and $\mathcal{O}(1/k)$ read
\begin{align}
\begin{split}
&\b^{a\hat{b}}=\frac{1}{2\sqrt{k_1k_2}}\mathcal{N}'_{ac}{}^{\hat{d}}\mathcal{N}'^{(T)}_{\hat{b}\hat{d}}{}^{c}+\mathcal{O}(\tilde{\l}^2), \\ &\b^{a\tilde{b}}=\frac{1}{2\sqrt{k_1k_2}}\mathcal{N}'_{ac}{}^{\hat{d}}\tilde{\l}^T_{\tilde{b}e}\l^T_{\hat{d}l}\tilde{g}^{fc}f_{elf}+\mathcal{O}(\tilde{\l}^2), \\
&\mathcal{N}'_{ab}{}^{\hat{c}}=\mathcal{N}'_{ab}{}^{\hat{c}}(\l,\l_0^{-1})=(\l_{a\hat{e}}\l_{b\hat{d}}f_{\hat{e}\hat{d}\hat{f}}-\l_0^{-1}\l_{e\hat{f}}f_{abe})g^{\hat{f}\hat{c}}, \\
&\mathcal{N}'_{\hat{a}\hat{b}}{}^{(T)c}=\mathcal{N}'_{\hat{a}\hat{b}}{}^{(T)c}(\l^T,\l_0)=(\l^T_{\hat{a}e}\l^T_{\hat{b}d}f_{edf}-\l_0\l^T_{\hat{e}f}f_{\hat{a}\hat{b}\hat{e}})\tilde{g}^{fc}, \label{beta 1st order}
\end{split}
\end{align}
where the components of $g$ and $\tilde{g}$ are the ones of (\ref{1st order}).
%The hats and tildes on the indices here are just for notational convenience and do not play any actual role since $\b^{a\hat{b}}=(\b_\l)^{ab}$ and $\b^{a\tilde{b}}=(\b_{\tilde{\l}})^{ab}$.
 As expected, $\b^{a\hat{b}}$ has the same form as (2.11) of \cite{Sagkrioti:2018rwg}, corresponding to the $\b$-function of the doubly deformed asymmetric model of \eqref{action2}.\\
\\
In order to compute the anomalous dimensions of the two composite operators we will need the covariant derivative of the two $\b$-functions in the $\tilde{\l}=0$ limit.
We are interested in the anomalous dimension components $(\g_{J_{1+}J_{2-}})_{a\hat{b}}{}^{c\hat{d}}$ and $(\g_{J_{1+}J_{3-}})_{a\tilde{b}}{}^{c\tilde{d}}$ referring to the corresponding interactions in the Lagrangian. %The only non-zero components of the Christoffel symbol are given in Appendix \ref{appendix:Christoffel}.
% along with the partial derivatives needed for the computation of the covariant derivative.
Using \eqref{1st order} and \eqref{beta 1st order}, the corresponding partial derivatives at the $\tilde{\l}=0$ limit are
\begin{align}
\begin{split}
&\partial_{m\hat{n}}\b^{a\hat{b}}=\frac{1}{2\sqrt{k_1k_2}}\left(\mathcal{K}_{ac;m\hat{n}}{}^{\hat{d}}\mathcal{N}'^{(T)}_{\hat{b}\hat{d}}{}^{c} +\mathcal{N}'_{ac}{}^{\hat{d}}\mathcal{K}^{(T)}_{\hat{b}\hat{d};\hat{n}m}{}^c \right),\qq \partial_{m\hat{n}}\b^{a\tilde{b}}=0,\\
&\partial_{m\tilde{n}}\b^{a\hat{b}}=0,\qq \partial_{m\tilde{n}}\b^{a\tilde{b}}=\frac{\d_{\tilde{n}}{}^{\tilde{b}}}{2\sqrt{k_1k_2}}\mathcal{N}'_{ac}{}^{\hat{d}}\l^T_{\hat{d}l}\tilde{g}^{fc}f_{mlf}, \label{partialbeta}
\end{split}
\end{align}
where
\begin{align*}
\begin{split}
&\mathcal{K}_{ab;i\hat{j}}{}^{\hat{c}}=\mathcal{K}_{ab;i\hat{j}}{}^{\hat{c}}(\l,\l_0^{-1})=(\d_{ai}\l_{b\hat{d}}-\d_{bi}\l_{a\hat{d}})f_{\hat{j}\hat{d}\hat{f}}g^{\hat{f}\hat{c}}-\l_0^{-1}f_{abi}g^{\hat{j}\hat{c}}+\mathcal{N}'_{ab}{}^{\hat{d}}(\l_{i\hat{k}}\d_{\hat{j}\hat{d}}+\l_{i\hat{d}}\d_{\hat{j}\hat{k}})g^{\hat{k}\hat{c}},\\
&\mathcal{K}_{\hat{a}\hat{b};\hat{j}i}^{(T)}{}^c=\mathcal{K}_{\hat{a}\hat{b};\hat{j}i}{}^c(\l^T,\l_0)=(\d_{\hat{a}\hat{j}}\l^T_{\hat{b}d}-\d_{\hat{b}\hat{j}}\l^T_{\hat{a}d})f_{idf}\tilde{g}^{fc}-\l_0f_{\hat{a}\hat{b}\hat{j}}\tilde{g}^{ic}+\mathcal{N}'^{(T)}_{\hat{a}\hat{b}}{}^d(\l^T_{\hat{j}k}\d_{id}+\l^T_{\hat{j}d}\d_{ik})\tilde{g}^{kc}\,.
\end{split}
\end{align*}
\\
Then, from \eqref{andimgeneralwithGammasSubstitutedintheDecouplingLimit} by substituting \eqref{1st order} and \eqref{partialbeta} we find the anomalous dimension components that we are interested in, to be\footnote{The indices in the l.h.s. are up(down), while in the r.h.s. are down(up) for the same reason as in footnote 4.}
\small
\begin{align}
\g_{a\hat{b}}{}^{c\hat{d}}=\frac{1}{2\sqrt{k_1k_2}}\Bigg[&\mathcal{K}_{ce;a\hat{b}}{} ^{\hat{s}}\mathcal{N}'^{(T)}_{\hat{d}\hat{s}}{}^e+\mathcal{N}'_{ce}{}^{\hat{s}}\mathcal{K}^{(T)}_{\hat{d}\hat{s};\hat{b}a}{}^e+\Big(\d^c_{m}\d_{\hat{b}}^{\hat{d}}(\l g^{-1})_{a\hat{n}}+\d_a^c\d^{\hat{d}}_{\hat{n}}(\l g^{-1})_{m\hat{b}}\Big)\mathcal{N}'_{m e}{}^{\hat{s}}\mathcal{N}'^{(T)}_{\hat{n}\hat{s}}{}^e\nonumber\\
& +\tilde{g}^{am}g^{\hat{b}\hat{n}}\tilde{g}_{cp}g_{\hat{d}\hat{q}}\bigg(\mathcal{K}_{me;p\hat{q}}{}^{\hat{s}}\mathcal{N}'^{(T)}_{\hat{n}\hat{s}}{}^e+\mathcal{N}'_{me}{}^{\hat{s}}\mathcal{K}^{(T)}_{\hat{n}\hat{s};\hat{q}p}{}^e\bigg) \label{andimJ1J2}\\
&+\tilde{g}^{am}g^{\hat{b}\hat{n}}\tilde{g}_{cp}g_{\hat{d}\hat{q}}\bigg(\d^{m}_i\d^{\hat{n}}_{\hat{q}}(\l g^{-1})_{p\hat{j}}+\d^{m}_p\d^{\hat{n}}_{\hat{j}}(\l g^{-1})_{i\hat{q}}   \bigg)\mathcal{N}'_{ie}{}^{\hat{s}}\mathcal{N}'^{(T)}_{\hat{j}\hat{s}}{}^e \Bigg]\ ,\nonumber
\end{align}
\normalsize
which, after some algebra, coincides with eq. (3.5) of \cite{Sagkrioti:2018abh}, as well as with eq. (3.9) of the same reference in the diagonal limit $\l_{a\hat{b}}=\l\d_{ab}$,  corresponding to the anomalous dimension of the composite $J_{1+}J_{2-}$ operator as expected, and
\begin{align}
\begin{split}
\g_{a\tilde{b}}{}^{c\tilde{d}}=\frac{\d_{\tilde{b}}{}^{\tilde{d}}}{2\sqrt{k_1k_2}}\Bigg[&\mathcal{N}'_{ce}{}^{\hat{s}}\l^T_{\hat{s}l}\tilde{g}^{fe}f_{alf}+(\l g^{-1})_{a\hat{m}}\mathcal{N}'_{ce}{}^{\hat{s}}\mathcal{N}'^{(T)}_{\hat{m}\hat{s}}{}^e\phantom{000000000000000}\\
&+\tilde{g}^{am}\tilde{g}_{cp}\Big(\mathcal{N}'_{m e}{}^{\hat{s}}\l^T_{\hat{s}l}\tilde{g}^{fe}f_{plf}+ (\l g^{-1})_{p\hat{i}}\mathcal{N}'_{m e}{}^{\hat{s}}\mathcal{N}'^{(T)}_{\hat{i}\hat{s}}{}^e  \Big)\Bigg]. \label{andimJ1J3 4indices}
\end{split}
\end{align}
However, there are also three more non-zero components given by
\begin{align}
\begin{split}
&\g_{ab}{}^{cd}=\tilde{g}_{cp}\b^{k\hat{l}}\partial_{k\hat{l}}\big(\tilde{g}^{pa}g^{db}\big),\\
&\g_{\hat{a}\hat{b}}{}^{\hat{c}\hat{d}}=g_{\hat{d}\hat{q}}\b^{k\hat{l}}\partial_{k\hat{l}}\big(\tilde{g}^{\hat{c}\hat{a}}g^{\hat{q}\hat{b}}\big),\\
&\g_{\tilde{a}\hat{b}}{}^{\tilde{c}\hat{d}}=g_{\hat{d}\hat{q}}\b^{k\hat{l}}\partial_{k\hat{l}}\big(\tilde{g}^{\tilde{p}\tilde{a}}g^{\hat{q}\hat{b}}\big), \label{other non-zero eigenvalues}
\end{split}
\end{align}
where the $g$ and $\tilde{g}$ components are the ones in \eqref{1st order}, while all other components of $\g_{AB}{}^{CD}$ vanish in the $\tilde{\l}=0$ limit.\\ Some comments regarding equations \eqref{andimJ1J2}-\eqref{other non-zero eigenvalues} are in order.
We may now rename all possible pairs of indices as $(ab)=\mathcal{A}$, $(a\hat{b})=\mathcal{B}$, $(a\tilde{b})=\mathcal{C}$, $(\hat{a}b)=\mathcal{D}$, etc. such that a $9\times9$ matrix is formed with each of its entries being a ($\dim G\times \dim G$)-dimensional block. It becomes obvious now that all non-zero components belong to the diagonal of that matrix and hence, in the $\tilde{\l}=0$ limit, correspond to the "eigenvalues" of the corresponding composite operators. The ones we are interested in are $\g_{a\hat{b}}{}^{c\hat{d}}=(\g_{J_{1+}J_{2-}})_{a\hat{b}}{}^{c\hat{d}}$ and $\g_{a\tilde{b}}{}^{c\tilde{d}}=(\g_{J_{1+}J_{3-}})_{a\tilde{b}}{}^{c\tilde{d}}$ corresponding to $J_{1+}J_{2-}$ and $J_{1+}J_{3-}$ respectively.
Notice here that (\ref{andimJ1J3 4indices}) is written as a tensor product with the $G_{k_3}$ copy being decoupled. This result would normally correspond to the anomalous dimension of the $J_{1+}J_{3-}$ composite operator, but since we are in the $\tilde{\l}=0$ limit, $J_{3-}$ does not appear in the action and thus, does not acquire an anomalous dimension.\\
 Then, $(\g_{J_{1+}J_{3-}})_{a\tilde{b}}{}^{c\tilde{d}}=(\g_{J_{1+}})_a{}^c\d_{\tilde{b}}{}^{\tilde{d}}$, with
\begin{align}
\begin{split}
(\g_{J_{1+}})_a{}^c=\frac{1}{2\sqrt{k_1k_2}}\Bigg[&\mathcal{N}'_{ce}{}^{\hat{s}}\l^T_{\hat{s}l}\tilde{g}^{fe}f_{alf}+(\l g^{-1})_{a\hat{m}}\mathcal{N}'_{ce}{}^{\hat{s}}\mathcal{N}'^{(T)}_{\hat{m}\hat{s}}{}^e\phantom{000000000000000}\\
&+\tilde{g}^{am}\tilde{g}_{cp}\Big(\mathcal{N}'_{m e}{}^{\hat{s}}\l^T_{\hat{s}l}\tilde{g}^{fe}f_{plf}+ (\l g^{-1})_{p\hat{i}}\mathcal{N}'_{m e}{}^{\hat{s}}\mathcal{N}'^{(T)}_{\hat{i}\hat{s}}{}^e  \Big)\Bigg], \label{andimJ1J3}
\end{split}
\end{align}
where the $\mathcal{N}'$ are the ones in \eqref{beta 1st order}.
Furthermore, in the diagonal and isotropic limit $\l_{a\hat{b}}=\l\d_{ab}$
\begin{equation}
(\g_{J_{1+}})_a{}^c=\g_{a}{}^{c}=\frac{c_G}{k_2}\frac{\l^2(\l-\l_0^{-1})^2}{(1-\l^2)^3}\d_a{}^c\,, \label{andim J1 diagonal}
\end{equation}
which matches with eq. (2.9) of \cite{Georgiou:2016zyo} with $k_2=k_R$, corresponding to the anomalous dimension of the single chiral current $J_{1+}$ in the isotropic limit.	\\
\begin{comment}
Finally, (\ref{andimJ1J2}) coincides with (3.5) of \cite{Sagkrioti:2018abh}, as well as with (3.9) of the same reference in the diagonal limit $\l_{ab}=\l\d_{ab}$,  corresponding to the anomalous dimension of the composite $J_{1+}J_{2-}$ operator as expected.\\
\end{comment}
Finally, the anomalous dimension of the anti-chiral current $J_{2-}$ is obtained from \eqref{andimJ1J3} by replacing $\l$ with $\l^T$ and exchanging $k_1\leftrightarrow k_2$, as explained in the previous Section. This has also been extensively checked with the use of the additional auxiliary term method. However, we chose not to present the tedious calculation here, since it repeats exactly the steps analysed in the present Section, but with the use of the coupling matrix \eqref{L2}, yielding the expected result. We should clarify here that the correct result for $J_{2-}$ obtained this way, cannot be recovered from the last equation of \eqref{other non-zero eigenvalues}, for reasons explained in the following paragraph.
 \begin{comment}
The result reads
\begin{align}
\begin{split}
\g_{\tilde{a}\hat{b}}{}^{\tilde{c}\hat{d}}=\frac{\d_{\tilde{a}}{}^{\tilde{c}}}{2\sqrt{k_1k_2}}\Bigg[&\mathcal{N}'^{(T)}_{\hat{d}\hat{e}}{}^s\l_{s\hat{l}}g^{\hat{f}\hat{e}}f_{\hat{b}\hat{l}\hat{f}}+(\l^T \tilde{g}^{-1})_{\hat{b}m}\mathcal{N}'_{m s}{}^{\hat{e}}\mathcal{N}'^{(T)}_{\hat{d}\hat{e}}{}^s \\
&+g^{\hat{b}\hat{m}}g_{\hat{d}\hat{p}}\Big(\mathcal{N}'^{(T)}_{\hat{m}\hat{e}}{}^s\l_{s\hat{l}}g^{\hat{f}\hat{e}} f_{\hat{p}\hat{l}\hat{f}}+(\l^T \tilde{g}^{-1})_{\hat{p}i}\mathcal{N}'_{is}{}^{\hat{e}}\mathcal{N}'^{(T)}_{\hat{m}\hat{e}}{}^s   \Big)  \Bigg] \label{andimJ2J3rewriten}
\end{split}
\end{align}
and has been extensively checked with the use of the additional ghost term method, which indeed yields the result \eqref{andimJ2J3rewriten}. We chose to not present the tedious calculation here, since it repeats exactly the steps analyzed in the present Section, but with the use of the coupling matrix \eqref{L2}.
\end{comment}
\\
The other three non-zero "eigenvalues" of \eqref{other non-zero eigenvalues} correspond to a part of the anomalous dimensions of  $J_{1+}J_{1-}$, $J_{2+}J_{2-}$ and $J_{3+}J_{2-}$ operators respectively. These operators do not appear in the action \eqref{action3}, but they develop non-zero anomalous dimension as a contribution coming from the individual single currents $J_{1+}$ and $J_{2-}$, which are indeed present in the initial action (even in the $\tilde{\l}=0$ limit) and acquire anomalous dimensions.
However, these are not the full anomalous dimensions of these three composite operators. % since \eqref{other non-zero eigenvalues} only includes the contribution
%of $J_{1-}$, $J_{2+}$
%coming from the interaction of $J_{1+}J_{2-}$ with itself\footnote{The contribution in the anomalous dimension of \eqref{other non-zero eigenvalues} comes only from the interaction of $J_{1+}J_{2-}$ with itself, and not with $J_{1+}J_{3-}$, since, in the $\tilde{\l}=0$ limit, the Zamolodchikov's metric is block-diagonal and only $\b^{a\hat{b}}$ is non-zero (see eq. \eqref{andimgeneralwithGammasSubstitutedintheDecouplingLimit}). The case of \eqref{andimJ1J3 4indices} differs, since partial derivatives with respect to $\tilde{\l}_{a\tilde{b}}$ are included.}.
 In order to obtain the full result, one needs to also include these operators in the initial action\footnote{The need to include the corresponding composite operator in the action is further supported by the fact that eq. \eqref{andimgeneral} has been proved to provide the anomalous dimensions of operators driving a theory away from a CFT \cite{Kutasov:1989dt}.}, similarly to the case of the auxiliary $J_{1+}J_{3-}$ operator,  and then take the limit where their coupling goes to zero\footnote{This procedure should be performed independently for each composite operator, in order to avoid the possible mixing of the anomalous dimension matrix due to the emergence of new directions in the RG flow. For example, if the operator $J_{3+}J_{2-}$ is added to \eqref{action2}, then the only two non-zero components of $\b^{AB}$ should be $\b^{a\hat{b}}$ and $\b^{\tilde{a}\hat{b}}$. Otherwise, the procedure should be modified in order to include diffeomorphisms and the resulting anomalous dimension matrix might not be block-diagonal, implying mixing between operators belonging in different blocks.}. Then, an additional contribution to the anomalous dimension of the auxiliary $J_{i+}J_{j-}$ will appear %due to its interaction with itself. This contribution appears 
 as the partial derivative terms of the corresponding $\b$-function in \eqref{andimgeneralwithGammasSubstituted} and in the limit where the auxiliary coupling goes to zero, survives and contributes only to the anomalous dimension of the corresponding operator without affecting the others. This happens because in the decoupling limit, in which we recover the $\l$-model \eqref{action2}, the Zamolodchikov's metric becomes block-diagonal (i.e. $g$ and $\tilde{g}$ are block-diagonal) and since the $\b$-function of the auxiliary coupling vanishes, there is no mixing in the anomalous dimensions. \\
Given the above arguments, it becomes clear now, why the correct result for the anomalous dimension of $J_{2-}$ cannot be obtained from the last equation of \eqref{other non-zero eigenvalues}.\\
\\
We should further note here, that our result does not change if we consider a $\L_{AB}$ matrix with all block interactions present in the initial action as auxiliary terms. The limit where all the auxiliary couplings go to zero, will again yield \eqref{andimJ1J3} as a final result, as a consequence of the block-diagonal form of the metric in this limit. In this case, no diffeomorphisms are needed, since all the block-entries of $\L_{AB}$ are considered non-zero and general, resulting to consistent RG flows.\\$\qq$\\

\section{The two couplings case using a coset and subgoup}
Consider again the model of \eqref{action2},
%\cite{Georgiou:2017jfi}
where we now split the group indices into subgroup $H\subset G$ and coset $G/H$ ones
% with a subgoup $H$ of the group G and the coset $G/H$.
and modify the split index $A$ notation of the previous Section, to $A=(a,\a)$, where the Latin/Greek indices stand for the subgroup/coset components. The hats and tildes are now omitted from the indices notation, since we work directly in the $\tilde{\l}=0$ limit, where all the results depend only on $\l_{a\hat{b}}\to\l_{AB}$. Then, for non-symmetric Einstein-spaces the following relations hold \cite{Forgacs:1985vp,Lust:1986ix}
\begin{align}
\begin{split}
&f_{ACD}f_{BCD}=c_G\d_{AB},\qq f_{acd}f_{bcd}=c_H\d_{ab},\qq f_{\a\g\d}f_{\b\g\d}=c_{G/H}\d_{\a\b},\\
&f_{a\g\d}f_{b\g\d}=(c_G-c_H)\d_{ab},\qq f_{\a\g c}f_{\b\g c}=\frac{1}{2}(c_G-c_{G/H})\d_{\a\b},\qq f_{ab\g}=0, \label{identities}
\end{split}
\end{align}
where $c_{G}$, $c_{H}$ are the eigenvalues of the quadratic Casimir operator in the adjoint representation for the group $G$ and the subgroup $H$ respectively, while $c_{G/H}$ is just a constant. For symmetric spaces $f_{\a\b\g}=0$ and $c_{G/H}=0$.\\
We will now work in the $\tilde{\l}=0$ limit of \eqref{action3} where the action \eqref{action2} is recovered. For a diagonal $\l_{AB}$ deformation, where the subgroup and the coset theory have different deformation parameters, the non-zero coupling components are $\l_{ab}=\l_H\d_{ab}$ and $\l_{\a\b}=\l_{G/H}\d_{\a\b}$. Our goal is to compute the anomalous dimensions for the subgroup and coset single current components $J_{1+}^a$, $J_{1+}^{\a}$ of the $\l$-model \eqref{action2}, using the results of Section 3.1.2. %From \eqref{andimJ1J3 4indices} we then have
%\begin{align}
%\begin{split}
%\g_{AB}{}^{CD}=\frac{\d_{B}{}^{D}}{2\sqrt{k_1k_2}} \g_A{}^C, \quad A=(a,\a),
%\end{split}
%\end{align}
%where, 
We can re-express \eqref{andimJ1J3} in terms of the $\b$-functions as
\begin{align}
\begin{split}
%\g_A{}^C=
(\g_{J_{1+}})_A{}^C=&\mathcal{N}'_{CE}{}^{S}\l^T_{SL}\tilde{g}^{FE}f_{ALF}+2\sqrt{k_1k_2}(\l g^{-1})_{AM}\b^{CM}\\
&+\tilde{g}^{AM}\tilde{g}_{CP}\Big(\mathcal{N}'_{ME}{}^{S}\l^T_{SL}\tilde{g}^{FE}f_{PLF}+2\sqrt{k_1k_2}(\l g^{-1})_{PI}\b^{MI} \Big), \label{coset+subgroup andim}
\end{split}
\end{align}
where the capital indices $A=(a,\a)$ now run in the subgroup and the coset respectively. The $\b$-functions $\b^{ab}=\b_{\l_H}\d^{ab}$ and $\b^{\a\b}=\b_{\l_{G/H}}\d^{\a\b}$ %$(\b^{ab},\b^{\a\b})$ are diagonal and 
have been computed before in \cite{Sagkrioti:2018rwg} to be
\begin{equation*}
\b_{\l_H} = -{(\l_H-\l_0)(\l_H-\l_0^{-1})\ov 2 \sqrt{k_1k_2}}
\left( c_H {\l^2_H \ov (1-\l_H^2)^2} +
(c_G-c_H) {\l_{G/H}^2 \ov (1-\l_{G/H}^2)^2} \right)\ ,
\end{equation*}
\begin{equation*}
\begin{split}
&\b_{\l_{G/H}} =
-{1\ov 2 \sqrt{k_1k_2}} \left(c_{G/H} {\l_{G/H}^2 (\l_{G/H}-\l_0)(\l_{G/H}-\l_0^{-1})\ov (1-\l_{G/H}^2)^2}
+ {c_G-c_{G/H}\ov 2}\times\right.
\\
& \left. \times {\l_{G/H} \ov (1-\l_{G/H}^2) (1-\l_H^2)}
\left((\l_0^{-1}-\l_H)(\l_0\l_H-\l_{G/H}^2)+ (\l_0-\l_H)(\l_0^{-1}\l_H-\l_{G/H}^2)\right) \right)\ .
\end{split}
\end{equation*}
These are also invariant under the transformation
\begin{equation}
\l_{G/H}\to\l_{G/H}^{-1}\ , \quad \l_H\to\l_H^{-1} \ ,\quad  k_{1,2}\to-k_{2,1} \ , \label{coset symmetry}
\end{equation}
while the different components of $\mathcal{N}'$ are %the ones in (3.2) of the same reference with $k_1\leftrightarrow k_2$, reading
given from \eqref{beta 1st order}, reading
\begin{equation}
\begin{split}
	& \cN'_{ab}{}^c(\l_H;\l_0^{-1}) = - {\l_H(\l_0^{-1}-\l_H)\ov 1-\l_H^2} f_{abc}\ ,\\
	&
	\cN'_{\a\b}{}^c(\l_H,\l_{G/H};\l_0^{-1}) = {\l_{G/H}^2-\l_0^{-1}\l_H\ov 1-\l_H^2} f_{\a\b c}\ ,
	\\
	&
	\cN'_{\a\b}{}^\g(\l_{G/H};\l_0^{-1}) = -{\l_{G/H}(\l_0^{-1}-\l_{G/H})\ov 1-\l_{G/H}^2} f_{\a\b\g}\ ,\\
	&\cN'_{\a b}{}^\g(\l_H,\l_{G/H};\l_0^{-1}) = -{\l_{G/H}(\l_0^{-1}-\l_H) \ov 1-\l_{G/H}^2} f_{\a b\g}\ ,
	\\
	&
	\cN'_{a\b}{}^\g(\l_H,\l_{G/H};\l_0^{-1}) = -{\l_{G/H}(\l_0^{-1}-\l_H) \ov 1-\l_{G/H}^2} f_{a\b\g}\ ,\\
	&\cN'_{a\b}{}^c = \cN'_{ab}{}^\g = \cN'_{\a b}{}^c= 0 \ .
\end{split}
\end{equation}
Using the fact that for the present choice of deformation matrix the $g$ and $\tilde{g}$ components of \eqref{1st order} are
\begin{align*}
&g_{ab}=\tilde{g}_{ab}=(1-\l_H^2)\d_{ab},\qq g_{\a\b}=\tilde{g}_{\a\b}=(1-\l_{G/H}^2)\d_{\a\b},\\
&g^{ab}=\tilde{g}^{ab}=(1-\l_H^2)^{-1}\d_{ab},\qq g^{\a\b}=\tilde{g}^{\a\b}=(1-\l_{G/H}^2)^{-1}\d_{\a\b},
\end{align*}
\eqref{coset+subgroup andim} yields the result $(\g_{J_{1+}})_a{}^c=(\g_{J^H_{1+}})\d_a{}^c$, with
\small
\begin{align}
\g_{J^H_{1+}}=\frac{(\l_0^{-1}-\l_H)^2}{k_2(1-\l_{G/H}^2)^2(1-\l_H^2)^3}\Bigg[c_G\l_{G/H}^2(1-\l_H^2)^2
-c_H(\l^2_{G/H}-\l_H^2)(1-\l_{G/H}^2\l_H^2)\Bigg] \label{andim subgroup}
\end{align}
\normalsize
for the anomalous dimension of the subgroup current component $J_{1+}^a$,
\newpage
and $(\g_{J_{1+}})_\a{}^\g=(\g_{J^{G/H}_{1+}})\d_\a{}^\g$, with
\begin{align}
\begin{split}
\g_{J^{G/H}_{1+}}=&\frac{1}{2\sqrt{k_1k_2}}\frac{1}{(1-\l_{G/H}^2)^3(1-\l_H^2)}\Bigg[c_G(1-\l_{G/H}^2)\times\\
&\times \Bigg(\l_0^{-1}(\l_{G/H}^2+\l_H^2)+\l_{G/H}^2\Big(\l_0(\l_{G/H}^2+\l_H^2)-4\l_H\Big)\Bigg)\\
&+c_{G/H}(\l_{G/H}-\l_H)\Bigg(\l_0^{-1}(1+\l_{G/H}^2)(\l_{G/H}+\l_H)\\
&+\l_{G/H}^2\Big(\l_0(1+\l_{G/H}^2)(\l_{G/H}+\l_H)-4(1+\l_H\l_{G/H})\Big)\Bigg)\Bigg] \label{andim coset}
\end{split}
\end{align}
for the anomalous dimension of the coset current component $J_{1+}^\a$, while $(\g_{J_{1+}})_a{}^\g=(\g_{J_{1+}})_\a{}^c=0$.\\
%\subsubsection*{Important limits}
Notice that both \eqref{andim subgroup} and \eqref{andim coset} are invariant under the symmetry \eqref{coset symmetry} and in the $\l_{G/H}=\l_H=\l$ limit are identical to \eqref{andim J1 diagonal} as expected.
Furthermore, in the $\l_{G/H}=0$ limit, \eqref{andim subgroup} consistently reduces to \eqref{andim J1 diagonal} with $c_G$ and $\l$ replaced by $c_H$ and $\l_H$ respectively, providing the anomalous dimension of the single current operator of a $\l_H$-deformed theory with group $G$, where only the subgroup theory is deformed. Moreover, by exchanging $k_1$ with $k_2$ in \eqref{andim subgroup} and \eqref{andim coset}, we obtain the anomalous dimension of $J_{2-}^a$ and $J^\a_{2-}$ subgroup and coset current components. Finally, in the equal levels limit with $\l_0=1$ we can obtain the anomalous dimension $\g_{J_+^H}=\g_{J_-^H}$ and $\g_{J_+^{G/H}}=\g_{J_-^{G/H}}$ for the subgroup and coset current components of the simply $\l$-deformed symmetric model.
\\
	We should note here that the coset limit defined by taking equal levels and $\l_H\to1$,  is not well defined in \eqref{andim coset}. This means that we cannot recover the anomalous dimension of a parafermion by performing a limiting procedure to the anomalous dimension of the single current. This is due to the non-local phase that is present for a parafermion but absent for a current. However, the $\l_H\to1$ limit is well defined for the subgroup current component \eqref{andim subgroup} for the case of an Abelian subgroup $H$ ($c_H=0$). %\\The parafermionic limit of $\l_H=1$ is not well defined in \eqref{andim coset}. More specifically, it goes to infinity as \begin{math}
	%\frac{c_G-c_{G/H}}{4k(1-\l_H)}\,.
	%\end{math}

\section{Two examples with SU(2)}
In the present Section we present two examples for $G=SU(2)$. The first one is for a diagonal, fully anisotropic deformation matrix, while the second one is for a general non-diagonal matrix where, as a sub-case, we consider the deformation matrix of \cite{Sfetsos:2015nya}. These two examples have proved to be very enlightening regarding the results obtained through the application of our method, while the discussion at the end of Subsection 5.1 can also be considered as a complement to the result found in Section 3 of \cite{Sagkrioti:2018abh}.
\subsection{Diagonal and anisotropic case}
Let us now consider the SU(2) case with
%$\tilde{\l}=0$ limit and
 a diagonal but anisotropic $\l_{ab}$ matrix of the form
\begin{equation}
\l_{a\hat{b}}=\begin{pmatrix}
\l_1 && 0 && 0\\
0 && \l_2 && 0\\
0 && 0 && \l_3
\end{pmatrix}\ , \label{l1}
\end{equation}
with the following $\b$-functions
\begin{equation}
\b_{\l_1}=\frac{(\l_0+\l_0^{-1})\l_1(\l_2^2+\l_3^2)-2\l_2\l_3(1+\l_1^2)}{\sqrt{k_1k_2}(1-\l_2^2)(1-\l_3^2)},\qq \text{and cyclic in 1,2,3}\ .
\end{equation}
For the above matrix, (\ref{andimJ1J3}) yields a diagonal anomalous dimension matrix for the single currents, with entries
\begin{equation}
%\g_{1}{}^{1}=
(\g_{J_{1+}})_{1}{}^{1}=-\frac{2}{\sqrt{k_1k_2}}\frac{4\l_1\l_2\l_3-(\l_0^{-1}+\l_0\l_1^2)(\l_2^2+\l_3^2)}{(1-\l_1^2)(1-\l_2^2)(1-\l_3^2)} \label{su(2) diagonal}
\end{equation}
and cyclic in 1,2,3.
 In the $\l_1=\l_2=\l_3=\l$ limit, \eqref{su(2) diagonal} reduces to \eqref{andim J1 diagonal} with $c_G=4$, while for the equal levels $k_1=k_2$, in the isotropic limit we recover the result of (3.21) in \cite{Georgiou:2016iom}, again with $c_G=4$, corresponding to the anomalous dimension of a single current for the $SU(2)$ group. Furthermore, the components given above are invariant under the duality-type transformation \eqref{generalised symmetry} with $(k_1\to-k_2,k_2\to-k_1,\l_i\to1/\l_i)$, as expected.\\
  Notice here also that for $\l_1=\l_2=\l$ \eqref{su(2) diagonal} coincides with the $(\g_{J_{1+}})_1{}^1$ and $(\g_{J_{1+}})_2{}^2$ components of \eqref{andim coset} with $c_G=4$, $c_H=0$ and $c_{G/H}=0$, corresponding to the $SU(2)/U(1)$ symmetric coset. However, the $SU(2)/U(1)$ coset limit defined by taking equal levels and $\l_3\to1$, $\l_1=\l_2=\l$ (where now $\l_3=\l_H$ corresponds to the subgroup deformation) is not well defined here for the reasons explained in the previous Section, and thus, the corresponding result of \cite{freefieldscoset} cannot be recovered via a limiting procedure. \\
 %This means that we cannot recover the anomalous dimension of an $SU(2)/U(1)$ parafermion by performing a limiting procedure to the anomalous dimension of the single current. This is due to the non-local phase that is present for a parafermion while it is absent for a current.
 Moreover, in the IR fixed point ($\l_1=\l_2=\l_3=\l_0$), \eqref{su(2) diagonal} equals with \begin{math}
 	\frac{4}{k_2-k_1}
 \end{math}, while the anomalous dimension of the anti-chiral current is zero, recovering the results of \cite{Georgiou:2016zyo}.
\\
\\
 Finally, we can obtain from \eqref{andimJ1J2} the anomalous dimension matrix for the composite operator $J_{1+}J_{2-}$, with non-zero components
 \begin{align}
 \begin{split}
 &\g_{1\hat{1}}{}^{1\hat{1}}=-\frac{2}{\sqrt{k_1k_2}}\frac{8\l_1\l_2\l_3-(\l_0+\l_0^{-1})(1+\l_1^2)(\l_2^2+\l_3^2)}{(1-\l_1^2)(1-\l_2^2)(1-\l_3^2)}\,, \\
 &\g_{1\hat{1}}{}^{2\hat{2}}=-\frac{4}{\sqrt{k_1k_2}}\frac{\l_3(1+\l_1^2)(1+\l_2^2)-(\l_0+\l_0^{-1})\l_1\l_2(1+\l_3^2)}{(1-\l_1^2)^2(1-\l_3^2)},\\
 &\g_{1\hat{2}}{}^{1\hat{2}}=\g_{2\hat{1}}{}^{2\hat{1}}=\frac{2}{\sqrt{k_1k_2}}\frac{(\l_0+\l_0^{-1})(\l_1^2\l_2^2+\l_3^2)-4\l_1\l_2\l_3}{(1-\l_1^2)(1-\l_2^2)(1-\l_3^2)},\\
 &\g_{1\hat{2}}{}^{2\hat{1}}=\g_{2\hat{1}}{}^{1\hat{2}}= \frac{2}{\sqrt{k_1k_2}}\frac{2\l_3(1+\l_1^2\l_2^2)-(\l_0+\l_0^{-1})\l_1\l_2(1+\l_3^2)}{(1-\l_1^2)(1-\l_2^2)(1-\l_3^2)}\,, \label{su(2) diag composite 9x9}
 \end{split}
 \end{align}
 and cyclic in 1,2,3. For the case of the composite operator, the $SU(2)/U(1)$ symmetric coset limit can be recovered for $k_1=k_2$, by sending the coupling corresponding to the deformation of the $U(1)$ subgroup $\l_3\to1$ and take $\l_1=\l_2=\l$. Then, as expected, one of the well-defined eigenvalues of the above $9\times9$ matrix is identical with eq. (4.34) of \cite{freefieldscoset}, yielding
 	\begin{equation}
 	\g_{J_{1+}J_{2-}}^{SU(2)/U(1)}=-\frac{2}{k}\frac{1+\l^2}{1-\l^2}\,. \label{CosetcompositeAnDimSU(2)/U(1)}
 	\end{equation}
The reason why the above limit is well defined in this case, is because in the composite operator the non-local parafermionic phase is cancelled out due to the equal contribution of a holomorphic and an anti-holomorphic parafermion. %, while all other components are zero.
\\It is worth noting here that the first two results of \eqref{su(2) diag composite 9x9}, along with their cyclic permutations, can also be reproduced by \eqref{andimgeneralwithGammasSubstituted}, using the reduced $3\times 3$ couplings' space metric, with line element
\begin{equation*}
(ds)^2=G_{a\hat{b}|c\hat{d}}\text{d}\l_{a\hat{b}}\text{d}\l_{c\hat{d}}=\frac{(\text{d}\l_{1})^2}{(1-\l_{1}^2)^2}+\frac{(\text{d}\l_{2})^2}{(1-\l_2^2)^2}+\frac{(\text{d}\l_{3})^2}{(1-\l_{3}^2)^2}\,,
\end{equation*}
instead of the $9\times 9$ tensor product metric of \eqref{Zamolo metric definition}. In this reduced space we acquire a $3\times3$ anomalous dimension matrix involving only the operators that are present in the action, i.e. $J_{1+}^1J_{2-}^{\hat{1}}$, $J_{1+}^2J_{2-}^{\hat{2}}$, $J_{1+}^3J_{2-}^{\hat{3}}$, where the upper indices denote the corresponding su(2) elements with couplings $\l_1,\l_2,\l_3$ respectively. On the contrary, in the full space with the $9\times9$ metric, additional elements occur, corresponding to the full spectrum of $J_{1+}^aJ_{2-}^{\hat{b}}$ composite operators, with the indices $a,\hat{b}$ running again in the adjoint representation of the su(2).
Of course, if we diagonalize both anomalous dimension matrices, the three eigenvalues of the reduced one, appear inside the set of eigenvalues of the $9\times9$ matrix\footnote{The well defined eigenvalue \eqref{CosetcompositeAnDimSU(2)/U(1)} for the $SU(2)/U(1)$ coset of the $9\times9$ anomalous dimension matrix, is also present in the set of eigenvalues of the $3\times3$ reduced one, as expected.}. The reason why in the complete space we are able to recover the anomalous dimension components for operators that are not present in the action, is because the result \eqref{andimJ1J2} has been derived by considering all the entries of the coupling matrix $\l_{a\hat{b}}$ (and also of the $\b^{a\hat{b}}$) non-zero (i.e. in the initial action all interactions were present). This differs from the case where the results \eqref{andimJ1J2}-\eqref{other non-zero eigenvalues} are produced, since the $\L_{AB}$ matrix and the corresponding $\b^{AB}$ have only two non-zero entries and the contribution from partial derivatives on these $\b$s with respect to all zero $\L$-entries has not been included in the calculations. This is the reason why \eqref{other non-zero eigenvalues} corresponds only to a part of the anomalous dimensions of operators that are not present in the action with deformation term $\mathcal{J}_{1+}^A\L_{AB}\mathcal{J}_{2-}^B$ (with the capital indices $A=(a,\hat{a},\tilde{a})$ running in the three copies of the group $G$ and the matrix $\L$ of \eqref{L1}), while the additional components of \eqref{su(2) diag composite 9x9} correspond to the full anomalous dimensions of operators not present in the action with deformation term $J_{1+}^a\l_{a\hat{b}}J_{2-}^{\hat{b}}$  (with the indices now running in the adjoint of su(2) and the matrix $\l$ of \eqref{l1}).\\
We should further note here, that for cases, like the one in the following example, where two or more entries of the $\l_{a\hat{b}}$ matrix are related to each other, the comparison of results between the reduced and the full couplings' space might not be immediate. This happens because in the reduced space, the composite operators will differ from the ones in the complete one and so, the directions of the different flows in the two spaces cannot be straightforwardly compared.\\
\\
Finally, there is a further subtlety here regarding the contribution coming from the covariant derivatives with respect to couplings that have been set to zero, that is worth being analysed. This issue clearly appears only when one uses equation \eqref{andimJ1J2} in order to obtain the anomalous dimension matrix for the composite operators $J_{1+}^aJ_{2-}^b$ in the full space of couplings. As has already been discussed, in that case additional elements appear, corresponding to the anomalous dimension of composite operators whose couplings are zero in the action (see eq. \eqref{su(2) diag composite 9x9}). Then, one could ask why the contribution from these couplings does not affect the results for the operators existing in the action. The answer is that it does, and one should be very careful in the interpretation of these results. The only cases where the aforementioned results are safe to be used, are those where the $9\times9$ metric is block-diagonal (i.e. the $g$ and $\tilde{g}$ are block-diagonal) and the $\b$-functions of the couplings whose limit to zero has been taken, are also zero (i.e. the zero limit is compatible with the RG flows). This ensures that all possible contribution coming from the partial derivatives (acting both on the $\b$-functions and the Zamolodchikov's metric) with respect to couplings set to zero, is multiplied by the corresponding (zero) $\b$-function and thus does not contribute. This means that the $\g_{i\hat{j}}{}^{m\hat{n}}$ component will only acquire contribution from the $\partial_{i\hat{j}}\b^{m\hat{n}}$ and $\b^{k\hat{l}}\partial_{k\hat{l}}\big(\tilde{g}^{pa}g^{\hat{q}\hat{b}}\big)$ terms, with the $\partial_{i\hat{j}}$ and the $\b^{a\hat{b}}$ corresponding only to existing in the action couplings now. This way, we can acquire the anomalous dimensions for operators that are not present in the action and in the same time ensure that these operators will not affect the result for the ones that are indeed involved in the action. Both the SU(2) examples presented in this work have a diagonal Zamolodchikov's metric and the limit of the zero $\l$-entries is compatible with the RG flows. Thus, our results for the composite operators in both examples are valid. \\
The above issue is not present in the derivation of equation \eqref{andimJ1J3} (and equivalently of \eqref{andimJ1J2}-\eqref{other non-zero eigenvalues}), since
 in the limit where all the auxiliary block-entries of $\L_{AB}$ are taken to zero,
  the $9\times 9$ metric is block-diagonal (see eq. \eqref{1st order}).
  Given that this metric appears in front of the partial derivatives in \eqref{andimgeneralwithGammasSubstituted}, it can be directly replaced there, by its block-diagonal form
  in the decoupling limit (i.e. all $\L_{AB}$ entries equal to zero except from $\l_{a\hat{b}}$), and the expression of $\g_{a\tilde{b}}{}^{c\tilde{d}}$ is 
\begin{equation*}
\g_{a\tilde{b}}{}^{c\tilde{d}}=\partial_{a\tilde{b}}\b^{c\tilde{d}}+\tilde{g}_{cp}g_{\tilde{d}\tilde{q}}\Big(\tilde{g}^{am}g^{\tilde{b}\tilde{n}}\partial_{p\tilde{q}}\b^{m\tilde{n}}+\beta^{k\hat{l}}\partial_{k\hat{l}}\big(\tilde{g}^{pa}g^{\tilde{q}\tilde{b}}\big)\Big)\,.
\end{equation*}
It is clear now, that derivatives  corresponding to $\l_{a\hat{b}}$ do not cause any problem since the partial derivatives with respect to the couplings inside this block, are multiplied by the corresponding $\b^{a\hat{b}}$ (which is non-zero only for couplings of operators that are present in the action, in a consistent zero limit of some $\l_{a\hat{b}}$-entries)\footnote{Derivatives with respect to $\l_{a\tilde{b}}$ (or any other auxiliary coupling matrix) are not an issue, since we have the freedom to choose all of its (their) entries to be non-zero. Therefore, the partial derivatives with respect to them are always well defined.}. \\Thus, the result \eqref{andimJ1J3} for the anomalous dimension matrix of the single current operator $J_{1+}$ is valid for any choice of $\l_{a\hat{b}}$ matrix.
%(for example, a term of the form $\nabla_{\hat{a}\hat{b}}\b^{m\hat{n}}$ or $\nabla_{a\hat{b}}\b^{\hat{m}\hat{n}}$ does not appear inside $\g_{a\tilde{b}}{}^{c\tilde{d}}$).
%This result  becomes diagonal in the IR fixed point, with entries \begin{math}
%\frac{4}{k_2-k_1}
%\end{math}, being in agreement with the CFT property $\g_{J_{1+}J_{2-}}=\g_{J_{1+}}+\g_{J_{2-}}$ and for equal levels reduces to the results of \cite{Georgiou:2015nka}.

\subsection{Non-diagonal case}
 We can now go beyond the diagonal case and consider a matrix of the form % \cite{Sfetsos:2015nya}

 \begin{equation}
 \l_{a\hat{b}}=\begin{pmatrix}
 \l_1 && \l_2 && 0\\
 -\l_2 && \l_1 && 0\\
 0 && 0 && \l_3
 \end{pmatrix}.\label{l in ex non-diagonal} %\l_1=\frac{\zeta^2(1+\l)^2+4\l}{\zeta^2(1+\l^2)+4},\quad \l_2=\frac{2\zeta(1-\l^2)}{\zeta^2(1+\l^2)+4},\quad \l_3=\l
 \end{equation}
For this choice of deformation matrix, no new RG flows are generated and so no diffeomorphisms are needed. Thus, the $\b$-functions are simply given by the first equation of \eqref{beta 1st order}
 \begin{align}
\begin{split}
&\b_{\l_1}=\l_1f(\l_1,\l_2,\l_3),\qq \b_{\l_2}=\l_2f(\l_1,\l_2,\l_3),\\
&f(\l_1,\l_2,\l_3)=\frac{(\l_0+\l_0^{-1})(\l_1^2+\l_2^2+\l_3^2)-2(1+\l_1^2+\l_2^2)\l_3}{\sqrt{k_1k_2}(1-\l_3^2)(1-\l_1^2-\l_2^2)},\\
&\b_{\l_3}=-\frac{2}{\sqrt{k_1k_2}}\frac{(\l_1^2+\l_2^2)(\l_0-\l_3)(\l_0^{-1}-\l_3)}{(1-\l_1^2-\l_2^2)^2}\,. \label{betas ex1}
\end{split}
\end{align}
% where $\zeta$ is another coupling parameter.\\
\begin{comment}
Since in all the preceding calculations for $\g_{AB}{}^{CD}$ the derivative $\partial_{IJ}\L_{AB}=\d_{IA}\d_{JB}$ was used considering all the entries independent from each other, one should in general proceed carefully with choices of deformation matrices as the one above. However, since in the computation of \eqref{andimJ1J3} only the derivatives with respect to $\tilde{\l}$ contributed in the $\tilde{\l}=0$ limit, we may proceed directly by just inserting the matrix \eqref{l in ex non-diagonal} in the result \eqref{andimJ1J3}. We should mention here that in general this should not be the case for the anomalous dimension of the $J_{1+}J_{2-}$ composite operator, since for the derivation of \eqref{andimJ1J2}, the partial derivative $\partial_{a\hat{b}}\b^{c\hat{d}}$ has been used, considering all the entries of the $\l$ matrix non-zero and independent from each other.
\end{comment}
 %\textcolor{red}{This issue is addressed at the end of this subsection.}
% More specifically, they should consider all five non-zero elements of the above matrix independent from each other during the computations, and only in the final result the limits $\l_{22}=\l_{11}=\l_1$ and $\l_{12}=-\l_{21}=\l_2$ will be considered.\\
%Then, (\ref{andimJ1J3}) gives the following anomalous dimension components for the single currents
The anomalous dimension components for the single currents can be computed from (\ref{andimJ1J3}) and read
\begin{align}
&(\g_{J_{1+}})_{1}{}^{1}=(\g_{J_{1+}})_{2}{}^{2}=\frac{2}{\sqrt{k_1k_2}}\frac{\l_0^{-1}(\l_1^2+\l_2^2+\l_3^2)\big(1+\l_0^2(\l_1^2+\l_2^2)\big)-4\l_3(\l_1^2+\l_2^2)}{(1-\l_1^2-\l_2^2)^2(1-\l_3^2)},\nonumber\\
&(\g_{J_{1+}})_{3}{}^{3}=\frac{4}{\sqrt{k_1k_2}}\frac{\l_0(\l_1^2+\l_2^2)(\l_0^{-1}-\l_3)^2}{(1-\l_1^2-\l_2^2)^2(1-\l_3^2)}\ , \label{andim ex2 su(2)}
\end{align}
while all non-diagonal components are zero\footnote{The diagonal form of the anomalous dimension matrix for this example is due to the fact that the off-diagonal entries of the matrix \eqref{l in ex non-diagonal} have opposite values. Of course, for any other choice of non-diagonal coupling matrix, the anomalous dimension is expected to be non-diagonal.\\
	 One such example is the $SU(2)$ case with the symmetric matrix \begin{math}
\l_{a\hat{b}}=\begin{pmatrix}
	\l_1 && \l_2 && 0\\
	\l_2 && \l_1 && 0\\
	0 && 0 && \l_3
	\end{pmatrix},
	\end{math} where again no diffeomorphisms are needed. Then, the $(\g_{J_{1+}})_a{}^c$ anomalous dimension matrix is non-diagonal, yet each of its entries is again invariant under the transformation \eqref{generalised symmetry}.}.
Notice here that the $\l_2=0$ and $\l_3=\l_1=\l$ limit is consistent, and eq. \eqref{andim J1 diagonal} is recovered from \eqref{andim ex2 su(2)} as expected. Once again, the result \eqref{andim ex2 su(2)} is invariant under the symmetry \eqref{generalised symmetry}, involving the inversion of the matrix $\l_{a\hat{b}}$ and the exchange of $k_{1,2}\leftrightarrow -k_{2,1}$.
 \\
 \\
 If we now further reduce the number of independent directions in the flow of $\l_{a\hat{b}}$, as in the integrable case of \cite{Sfetsos:2015nya} where \begin{equation}\l_1=\frac{\zeta^2(1+\l)^2+4\l}{\zeta^2(1+\l)^2+4},\quad \l_2=\frac{2\zeta(1-\l^2)}{\zeta^2(1+\l)^2+4},\quad \l_3=\l \ , \end{equation}
  diffeomorphisms are indeed needed in order to consistently reduce the three-dimensional space of couplings $(\l_1,\l_2,\l_3)$ to a two-dimensional one $(\l,\zeta)$.\\
  According to Appendix A, we find the diffeomorphism $\hat{\zeta}'(\l,\zeta;\l_0)=\big(0,0,\hat{\zeta}'^3(\l,\zeta;\l_0)\big)$ where
  \small
  \begin{equation}
  \hat{\zeta}'^3(\l,\zeta;\l_0)=\zeta\frac{(1+\l_0^{-2})\big(4\l+(1+\l)^2\zeta^2\big)-2\l_0^{-1}\big(4(1-\l+\l^2)+(1+\l)^2\zeta^2\big)}{4\sqrt{2}(\l-\l_0^{-1})(1-\l)^2}\,. \label{zeta3difeo}
  \end{equation}
  \normalsize
  This model corresponds to a generalization of the one analysed in \cite{Sfetsos:2015nya}, with different Kac-Moody levels.\\
Then, the $\b$-functions for the new couplings $\l$ and $\zeta$ in the presence of the diffeomorphism \eqref{zeta3difeo} are
   \begin{align}
   \begin{split}
   &\b_\l=-\frac{(\l_0-\l)(\l_0^{-1}-\l)\big(4+(1+\l)^2\zeta^2\big)\big(4\l^2+(1+\l)^2\zeta^2\big)}{8\sqrt{k_1k_2}(1-\l^2)^2},\\
   &\b_\zeta=\frac{(2+\l_0+\l_0^{-1})\zeta \big(16\l^2+4(1+\l^2)(1+\l)^2\zeta^2+(1+\l)^4\zeta^4\big)}{16\sqrt{k_1k_2}(1-\l)(1+\l)^3}\ . \label{bl,bz}
   \end{split}
   \end{align}
 In this case the initial $\b_{\l_i}$, $i=1,2,3$ of \eqref{betas ex1} acquire an extra contribution of the form $\mathfrak{D}^{a\hat{b}}$ as given in \eqref{diffeos}. Then, combining \eqref{andimgeneralwithGammasSubstituted} with \eqref{diffeos contribution to andim general} and \eqref{term 4} in \eqref{andim dif}, the anomalous dimension matrix of $J_{1+}$ should be \eqref{andim ex2 su(2)} with an additional contribution of the form \begin{math}
 (\mathfrak{Dif})_{a\tilde{b}}{}^{c\tilde{d}}=\frac{1}{2\sqrt{k_1k_2}}\left(\d_p{}^a\d_c{}^m+\tilde{g}^{am}\tilde{g}_{cp}\right)\left(\l_0\l_{p\hat{s}}\mathcal{N}'_{mr}{}^{\hat{s}}-f_{mrp}\right)\hat{\zeta}'^r
 \end{math}. However, in the example at hand, the contribution of diffeomorphisms in the anomalous dimensions is zero and so, \eqref{andim ex2 su(2)} remain unchanged. Its components, in terms of the couplings $\l$ and $\zeta$ then read
 % Finally, in terms of $\l$ and $\zeta$, the \eqref{andim ex2 su(2)} components become
 \begin{align}
 &(\g_{J_{1+}})_{1}{}^{1}= (\g_{J_{1+}})_{2}{}^{2}=\frac{1}{8\sqrt{k_1k_2}(1-\l^2)^3}\Bigg[\big(4\l^2+(1+\l)^2\zeta^2\big)\Big((1+\l)^2\zeta^2\big(\l_0(1+\l^2)-4\l\big)\nonumber\\
 &\phantom{000000000}-8\l(2-\l_0\l)\Big)+\l_0^{-1}\big(32\l^2+4(1+\l)^2(1+3\l^2)\zeta^2+(1+\l)^4(1+\l^2)\zeta^4\big)\Bigg]\,\nonumber\\
 %&\g_{1\tilde{b}}{}^{2\tilde{d}}=-\g_{2\tilde{b}}{}^{1\tilde{d}}=-\frac{4(1-\l_0\l^3)+(1+\l)^2(1-\l_0\l)\zeta^2}{4\sqrt{2k_1k_2}(1-\l^2)}\hat{\zeta}'^3(\l,\zeta;\l_0)\ ,\label{andim with l and z couplings}\\
 &(\g_{J_{1+}})_{3}{}^{3}=\frac{\l_0(\l-\l_0^{-1})^2\big(4+(1+\l)^2\zeta^2\big)\big(4\l^2+(1+\l)^2\zeta^2\big)}{4\sqrt{k_1k_2}(1-\l^2)^3}\  \label{andim with l and z couplings}
 \end{align}
 and are invariant under the transformation
 \begin{equation*}
 \l\to\l^{-1},\qq \zeta\to-\zeta,\qq k_1\to-k_2,\qq k_2\to-k_1,
 \end{equation*}
 which is the manifestation of the duality-type symmetry in the reduced couplings' space $(\l,\zeta)$.
\begin{comment}
 Finally, we can also compute the anomalous dimension of the composite operator $J_{1+}J_{2-}$ from \eqref{andimJ1J2}, which has the following components
 \begin{align}
 &\g_{11}{}^{11}=\g_{22}{}^{22}=\frac{2}{\sqrt{k_1k_2}}\frac{(\l_0+\l_0^{-1})\big(\l_1^2+\l_3^2+\l_1^2\l_3^2+(\l_1^2+\l_2^2)^2\big)-8\l_3\l_1^2-4\l_3\l_2^2}{(1-\l_1^2-\l_2^2)^2(1-\l_3^2)}    \,, \nonumber\\
 &\g_{11}{}^{22}=\g_{22}{}^{11}=\frac{2}{\sqrt{k_1k_2}}\frac{(\l_0+\l_0^{-1})(\l_2^2+2\l_1^2)(1+\l_3^2)-2\l_3(1+2\l_1^2+(\l_1^2+\l_2^2)^2)}{(1-\l_1^2-\l_2^2)^2(1-\l_3^2)}\,,\nonumber\\
 &\g_{11}{}^{33}=\g_{22}{}^{33}=-\frac{4}{\sqrt{k_1k_2}}\frac{\l_1(1+\l_1^2+\l_2^2)\big(1-(\l_0+\l_0^{-1})\l_3+\l_3^2\big)}{(1-\l_1^2-\l_2^2)^3}\,,\\
 &\g_{33}{}^{11}=\g_{33}{}^{22}=\frac{(1-\l_1^2-\l_2^2)^2}{(1-\l_3^2)^2}\g_1{}^3\,,\nonumber\\
 &\g_{33}{}^{33}=\frac{4}{\sqrt{k_1k_2}}\frac{(\l_1^2+\l_2^2)\big((\l_0+\l_0^{-1})(1+\l_3^2)-4\l_3\big)}{(1-\l_1^2-\l_2^2)^2(1-\l_3^2)} \,. \nonumber
 \end{align}
 We notice here that since eq. \eqref{andimJ1J2} has been derived for the case of a general matrix with all entries non-zero,
 A comment is in order.
\end{comment}
\\
\\
Finally, some comments regarding the anomalous dimension of the composite operators in terms of $\l_1,\l_2,\l_3$ are in order. It is possible to compute the anomalous dimension matrix of $J_{1+}J_{2-}$ both in the complete space using \eqref{andimJ1J2} directly, or in the reduced one by using the corresponding $3\times 3$ metric
\begin{equation*}
(ds)^2=
%G_{AB|CD}\text{d}\L_{AB}\text{d}\L_{CD}=
\frac{(\text{d}\l_{1})^2}{(1-\l_{1}^2-\l_2^2)^2}+\frac{(\text{d}\l_{2})^2}{(1-\l_{1}^2-\l_2^2)^2}+\frac{1}{2}\frac{(\text{d}\l_{3})^2}{(1-\l_{3}^2)^2}\,.
\end{equation*}
 However, as explained in the end of the previous Subsection, a comparison between the two results now would not be immediate. For the case at hand, with the deformation matrix \eqref{l in ex non-diagonal}, the operators that are present in the action, in the reduced space are collected to the following new ones: $\mathcal{O}^1=J_{1+}^1J_{2-}^{\hat{1}}+J_{1+}^2J_{2-}^{\hat{2}}$, $\mathcal{O}^2=J_{1+}^1J_{2-}^{\hat{2}}-J_{1+}^2J_{2-}^{\hat{1}}$ and
	$\mathcal{O}^3=J_{1+}^3J_{2-}^{\hat{3}}$ with couplings $\l_1,\l_2,\l_3$ respectively. It is now clear that the only anomalous dimension matrix element that can be immediately compared and should be equal between the two spaces is the one corresponding to the $J_{1+}^3J_{2-}^{\hat{3}}$ operator, which is indeed the case. Of course, the three eigenvalues of the reduced anomalous dimension matrix, are again present in those of the $9\times9$ matrix.
	% In this case, one should be also carefull about the extra components that are present in the $9\times9$ matrix, corresponding to the operators that are not involved in the action. For the previous example with the diagonal deformation matrix the elements corresponding o zero couplings in the action couldn't cause any issue.

 \section{Discussion and future directions}
  In this work we presented an easy way to compute exact expressions for the anomalous dimensions of the single current operators for a doubly deformed $\l$-model with general deformation matrix. Our method was derived for the case of a general semi-simple, compact group and was applied in the case of a general non-symmetric coset, corresponding to an Einstein space. Two examples for the $SU(2)$ case were also presented, for diagonal and non-diagonal anisotropic $\l$ matrices. Finally, the case with diffeomorphisms was discussed in Appendix A. All results are exact in the deformation parameter(s) up to order $\mathcal{O}(1/k)$ and in agreement with limit cases existing in the literature. \\
  The computation was based in a combination of two methods introduced in \cite{exact results} and uses only the couplings' space geometry data along with an extra auxiliary group interaction which decouples at a certain limit. There are three major advantages in this approach:
 Firstly, there is no need for any perturbative calculation, which would have been tedious or even impossible in the case of a general $\l$ matrix.
 Secondly, given the fact that the method presented here takes advantage of the target space geometry in the computation of the $\b$-functions, it further incorporates information regarding the diffeomorphisms on the target space.
 Thirdly, this method allows us to compute exact quantum quantities with respect to every auxiliary coupling added in the initial action.
 \\
 In this context, we considered one extra interaction term, computed the exact in both couplings $\b$-functions, and then took the limit where the auxiliary coupling goes to zero, in order to obtain the anomalous dimension of the single current in the original theory. However, we could have kept the $\tilde{\l}$ coupling finite and compute the exact, both in $\l$ and $\tilde{\l}$, result for the anomalous dimensions of the two composite operators in the new theory. Another idea is to perform further deformations of the original action by introducing different auxiliary groups and include self- and/or mutual-interactions. This idea is currently under consideration. Finally, the method analysed in this work could be used in order to easily obtain higher order results in the 1/k-expansion, still exact in the deformation(s) parameters.

 \section*{Acknowledgments}
 I would like to thank K. Sfetsos for his suggestions and very fruitful discussions and K. Siampos for the useful comments and feedback on the manuscript.\\
 This research is co-financed by Greece and the European Union (European Social Fund- ESF) through the Operational Programme "Human Resources Development, Education and Lifelong Learning" in the context of the project "Strengthening Human Resources Research Potential via Doctorate Research" (MIS-5000432), implemented by the State Scholarships Foundation (IKY).

\appendix
\section{Anomalous dimensions with diffeomorphisms}

Consider now that the choice of deformation matrix is such that new flows are generated during the RG flow. Then, diffeomorphism terms should also be considered in order to cancel the flows in the new directions. In this case, the most general form of the $\b$-function can be found in (A.5) of \cite{Sagkrioti:2018abh}, and in the context of this work reads
\begin{equation}
\frac{\text{d}\L_{AB}}{\text{d}t}=\frac{1}{2}\left(\mathcal{N}_{AC}{}^D\mathcal{N}^{(T)C}_{BD}+\mathcal{N}_{AC}{}^D g_{BD}\hat{\zeta}^C\right), \qq \hat{\zeta}^C=-\tilde{g}^{CE}\hat{\xi}_E, \label{beta diffeos}
\end{equation}
with $\hat{\zeta}$ and $\hat{\xi}$ being constants with respect to the space-time coordinates, but in general depending on the coupling parameters and the levels $k_{1,2}$.\\
The first term of \eqref{beta diffeos} corresponds to the usual $\b$-function components of \eqref{bl} and \eqref{blt} (or \eqref{beta 1st order}) found in the main text, while the second one corresponds to the additional contribution of diffeomorphisms. The exact in $\l$ and $\tilde{\l}$ diffeomorphism components that would be added in \eqref{bl} and \eqref{blt} are given from the above relation with the $g$ and $\mathcal{N}$ components of \eqref{l,g,gt,d,f,h} and \eqref{Ns} accordingly.  \\
\\
We will now compute the contribution of the extra term to the $\b$-functions \eqref{beta 1st order} and the anomalous dimensions \eqref{andimJ1J2} and \eqref{andimJ1J3} up to order $\mathcal{O}(1/k)$ and in the $\tilde{\l}=0$ limit. By defining
\begin{equation*}
\mathfrak{D}^{AB}=\frac{1}{2}\mathcal{N}_{AC}{}^Dg_{BD}\hat{\zeta}^C,
\end{equation*}
after some algebra, the following contributions to $\b^{a\hat{b}}$ and $\b^{a\tilde{b}}$ up to $\mathcal{O}(\tilde{\l})$ can be found
\begin{equation}
\mathfrak{D}^{a\hat{b}}=\frac{\l_0}{2\sqrt{k_1k_2}}\mathcal{N}'_{ac}{}^{\hat{d}}g_{\hat{b}\hat{d}}\hat{\zeta}'^c,\quad \mathfrak{D}^{a\tilde{b}}=-\frac{1}{2\sqrt{k_1k_2}}\tilde{\l}_{r \tilde{b}}f_{acr}\hat{\zeta}'^c,\qq \hat{\zeta}'^c=\sqrt{k_2}\hat{\zeta}^c, \label{diffeos}
\end{equation}
while $\mathfrak{D}^{ab}=\mathfrak{D}^{\hat{a}B}=\mathfrak{D}^{\tilde{a}B}=0$, $B=(b,\hat{b},\tilde{b})$. The expressions for $g$ and $\mathcal{N}'$ in \eqref{diffeos} are the ones in \eqref{1st order} and \eqref{beta 1st order}.\\
\\
The anomalous dimensions for the composite operators are again given by \eqref{andimgeneral}, but now with the $\b$-function of \eqref{beta diffeos}. The former, compared to the ones found in the main text, would have an additional contribution of the form
\begin{align}
\begin{split}
\big(\mathfrak{Dif}\big)_{AB}{}^{CD}&=\nabla_{AB}\mathfrak{D}^{CD}+\nabla^{CD}\mathfrak{D}_{AB}=\nabla_{AB}\mathfrak{D}^{CD}+G_{AB|MN}G^{CD|PQ}\nabla_{PQ}\mathfrak{D}^{MN}\\
&=\partial_{AB}\mathfrak{D}^{CD}+\tilde{g}_{CP}g_{DQ}\bigg(\tilde{g}^{AM}g^{BN}\partial_{PQ}\mathfrak{D}^{MN}+\mathfrak{D}^{k\hat{l}}\partial_{k\hat{l}}\big(\tilde{g}^{PA}g^{QB}\big)\bigg). \label{diffeos contribution to andim general}
\end{split}
\end{align}
Then \eqref{diffeos contribution to andim general} has the following non-zero components
\begin{align}
\begin{split}
&\big(\mathfrak{Dif}\big)_{ab}{}^{cd}=\tilde{g}_{cp}\mathfrak{D}^{k\hat{l}}\partial_{k\hat{l}}\big(\tilde{g}^{ap}g^{bd}\big)\,,\quad \\	&\big(\mathfrak{Dif}\big)_{\hat{a}\hat{b}}{}^{\hat{c}\hat{d}}=g_{\hat{d}\hat{q}}\mathfrak{D}^{k\hat{l}}\partial_{k\hat{l}}\big(\tilde{g}^{\hat{a}\hat{c}}g^{\hat{b}\hat{q}}\big)\,,\quad \\
&\big(\mathfrak{Dif}\big)_{\tilde{a}\hat{b}}{}^{\tilde{c}\hat{d}}=g_{\hat{d}\hat{q}}\mathfrak{D}^{k\hat{l}}\partial_{k\hat{l}}\big(\tilde{g}^{\tilde{c}\tilde{p}}g^{\hat{b}\hat{q}}\big)\,,
 \label{term 1}
\end{split}
\end{align}
where the $g,\tilde{g}$ and $\mathfrak{D}^{AB}$ components are the ones in \eqref{1st order} and \eqref{diffeos} respectively, along with
\begin{align}
\begin{split}
\big(\mathfrak{Dif}\big)_{a\hat{b}}{}^{c\hat{d}}=&\big(\d_p{}^a\d_{\hat{q}}{}^{\hat{b}}\d_c{}^m\d_{\hat{d}}{}^{\hat{n}}+\tilde{g}_{cp}g_{\hat{d}\hat{q}}\tilde{g}^{am}g^{\hat{b}\hat{n}}\big)\partial_{p\hat{q}}\mathfrak{D}^{m\hat{n}} \\
&+\tilde{g}_{cp}g_{\hat{d}\hat{q}}\mathfrak{D}^{k\hat{l}}\Big(\partial_{k\hat{l}}\tilde{g}^{ap}g^{\hat{b}\hat{q}}+\tilde{g}^{ap}\partial_{k\hat{l}}g^{\hat{b}\hat{q}} \Big)     \,,  \label{term 2}
\end{split}
\end{align}
\begin{equation}
\begin{split}
&\big(\mathfrak{Dif}\big)_{a\tilde{b}}{}^{c\hat{d}}=\partial_{a\tilde{b}}\mathfrak{D}^{c\hat{d}}=\frac{\l_0}{2\sqrt{k_1k_2}}\mathcal{N}'_{cr}{}^{\hat{s}}g_{\hat{d}\hat{s}}\partial_{a\tilde{b}}\hat{\zeta}'^r\,,\\
&\big(\mathfrak{Dif}\big)_{a\hat{b}}{}^{c\tilde{d}}=\frac{\l_0}{2\sqrt{k_1k_2}}\tilde{g}_{cp}g_{\tilde{d}\tilde{q}}\tilde{g}^{am}g^{\hat{b}\hat{n}}\mathcal{N}'_{mr}{}^{\hat{s}}g_{\hat{n}\hat{s}}\partial_{p\tilde{q}}\hat{\zeta}'^r \,, \label{term 3}
\end{split}
\end{equation}
\begin{equation}
\begin{split}
\big(\mathfrak{Dif}\big)_{a\tilde{b}}{}^{c\tilde{d}}=&\big(\d_p{}^a\d_{\tilde{q}}{}^{\tilde{b}}\d_c{}^m\d_{\tilde{d}}{}^{\tilde{n}}+\tilde{g}_{cp}g_{\tilde{d}\tilde{q}}\tilde{g}^{am}g^{\tilde{b}\tilde{n}}\big)\partial_{p\tilde{q}}\mathfrak{D}^{m\tilde{n}} \\
&+\tilde{g}_{cp}g_{\tilde{d}\tilde{q}}\mathfrak{D}^{k\hat{l}}\partial_{k\hat{l}}\tilde{g}^{ap}g^{\tilde{b}\tilde{q}}     \,, \label{term 4}
\end{split}
\end{equation}
where
\begin{align}
&\partial_{a\hat{b}}\mathfrak{D}^{c\hat{d}}=\frac{\l_0}{2\sqrt{k_1k_2}}\Bigg(\mathcal{K}_{cr;a\hat{b}}{}^{\hat{s}}(\l,\l_0^{-1})g_{\hat{d}\hat{s}}\hat{\zeta}'^r-\mathcal{N}'_{cr}{}^{\hat{s}}\big(\d_{\hat{b}\hat{d}}\l_{a\hat{s}}+\d_{\hat{s}\hat{b}}\l_{a\hat{d}}\big)\hat{\zeta}'^r+\mathcal{N}'_{cr}{}^{\hat{s}}g_{\hat{d}\hat{s}}\partial_{a\hat{b}}\hat{\zeta}'^r\Bigg)\,,\nonumber\\
&\partial_{a\tilde{b}}\mathfrak{D}^{c\tilde{d}}=-\frac{\d_{\tilde{b}}{}^{\tilde{d}}}{2\sqrt{k_1k_2}}f_{cra}\hat{\zeta}'^r\,, \nonumber\\ &\partial_{a\hat{b}}\tilde{g}^{cd}=\frac{\partial (1-\l\l^T)_{cd}^{-1}}{\partial\l_{a\hat{b}}}=\tilde{g}^{cm}\big(\l^T_{\hat{b}i}\d_{am}+\l_{m\hat{b}}\d_{ai}\big)\tilde{g}^{id}\,,\label{partialsDiff}\\
&\partial_{a\hat{b}}g^{\hat{c}\hat{d}}=\frac{\partial(1-\l^T\l)^{-1}}{\partial\l_{a\hat{b}}}=g^{\hat{c}\hat{m}}\big(\l_{a\hat{i}}\d_{\hat{b}\hat{m}}+\l^T_{\hat{m}a}\d_{\hat{b}\hat{i}}\big)g^{\hat{i}\hat{d}}\,. \nonumber
\end{align}
The $g,\tilde{g}$ and and the $\mathcal{N}'$ components in the above relations  are again given in \eqref{1st order}, \eqref{beta 1st order}, while the $\mathcal{K}$ is the one defined bellow \eqref{partialbeta}. Of course, we are interested only in the contribution to $\g_{a\hat{b}}{}^{c\hat{d}}$ and $\g_{a\tilde{b}}{}^{c\tilde{d}}$, i.e. eqs \eqref{term 2} and \eqref{term 4}, where in \eqref{term 4}, after substitution of \eqref{partialsDiff}, the third copy of the group decouples again.\\ Notice again that the contributions of \eqref{term 1}, \eqref{term 2} and \eqref{term 4} correspond again to block-diagonal elements of a $9\times9$ matrix, while the \eqref{term 3} contribution to $\g_{a\tilde{b}}{}^{c\hat{d}}$ and $\g_{a\hat{b}}{}^{c\tilde{d}}$ is off-diagonal. If $\hat{\zeta}'$ contained a linear in $\tilde{\l}$ term, then \eqref{term 3} would not vanish, implying a mixing of the two composite operators that are present in the action \eqref{action3}, which would remain even in the $\tilde{\l}=0$ limit. However, from the $\b^{a\hat{b}}$ of \eqref{beta 1st order}, it is obvious that the $\tilde{\l}$ matrix affects the flow of $\l_{a\hat{b}}$ at order $\tilde{\l}^2$ and higher. This implies that any diffeomorphism term, used to cancel the extra flows in the $\b^{a\hat{b}}$ block, will be of the form $\hat{\zeta}'(\l,\tilde{\l};\l_0)=\hat{\zeta}'_{(0)}(\l;\l_0)+\mathcal{O}(\tilde{\l}^2)$. Then, in the $\tilde{\l}=0$ limit, \eqref{term 3} vanishes and the aforementioned limit indeed acts as a decoupling limit.\\
Summarizing, the anomalous dimension tensor in the case where diffeomorphisms need to be included is
\begin{equation}
(\g_{AB}{}^{CD})_{\text{diff.}}=\g_{AB}{}^{CD}+\big(\mathfrak{Dif}\big)_{AB}{}^{CD}\,, \label{andim dif}
\end{equation}
where the components of $\g_{AB}{}^{CD}$ are the those in \eqref{andimJ1J2}, \eqref{andimJ1J3 4indices} and \eqref{other non-zero eigenvalues}, while the non-zero components of $\big(\mathfrak{Dif}\big)_{AB}{}^{CD}$ are those of \eqref{term 1}, \eqref{term 2} and \eqref{term 4}.
\\
\\
A final note here concerning diffeomorphisms is the following: The choice of $\l_{a\hat{b}}$ matrix can also affect the flows in the $\b^{a\tilde{b}}$ block and generate new directions. However, since the choice of $\tilde{\l}_{a\tilde{b}}$ does not affect our results in the decoupling limit, this issue is resolved by choosing a $\tilde{\l}$ matrix with all entries non-zero and independent from each other. Thus, it suffices for the diffeomorphism $\hat{\zeta}'$ to be such that only the $\b^{a\hat{b}}$ flows are corrected.

\def\baselinestretch{1.2}
\baselineskip 20 pt
\noindent

%%%%%%%%%%%%%%%

%\newpage

\end{document}